\documentclass[9pt,twocolumn,twoside]{osajnl}

\usepackage{amssymb,amsmath,stmaryrd,array}

\usepackage{subfig}

\makeatletter

\newcommand{\ket}[1]{\ensuremath{|{#1}\rangle}}
\newcommand{\bra}[1]{\ensuremath{\langle{#1}|}}
\newcommand{\sca}[2]{\ensuremath{\bigl({#1}\cdot{#2}\bigr)}}
\newcommand{\avr}[1]{\ensuremath{\langle{#1}\rangle}}

\newcommand{\cnj}[1]{{#1}^{\ast}}
\newcommand{\hcnj}[1]{{#1}^{\dagger}}

\newcommand{\Tr}{\mathop{\rm Tr}\nolimits}

 \newcommand{\vc}[1]{\mathbf{#1}}
 \newcommand{\mvc}[1]{\mathbf{#1}}
 \newcommand{\uvc}[1]{\hat{\mathbf{#1}}}
 
 \newcommand{\ind}[1]{\mathrm{#1}}

\newcommand{\dd}{\mathrm{d}}

 \newcommand{\e}{\mathrm{e}}

\makeatother

\journal{josab} 

\setboolean{shortarticle}{false} 

\ifthenelse{\boolean{shortarticle}}{\colorlet{color2}{color2b}}{\colorlet{color2}{color2}}

\title{
Algebraic approach to
electro-optic modulation
of light:\\  
Exactly solvable multimode quantum model 
}

\author[1]{George~P.~Miroshnichenko}
\author[1,*]{Alexei~D.~Kiselev}
\author[1]{Alexander~I.~Trifanov}
\author[1]{Artur~V.~Gleim}

\affil[1]{Saint Petersburg National Research University of Information Technologies,
 Mechanics and Optics (ITMO University),
 Kronverksky Prospekt 49,
 197101 Saint Petersburg, Russia}
\affil[*]{Corresponding author: alexei.d.kiselev@gmail.com}

\dates{Compiled \today}

\ociscodes{
(230.4140) Modulators;
(270.4180) Multiphoton processes; 
(270.5290) Photon counting; 
(270.5565) Quantum communications;
(190.4975) Parametric processes;
}

\doi{\url{http://dx.doi.org/10.1364/ao.XX.XXXXXX}}

\begin{abstract}
We theoretically study electro-optic light modulation
based on a quantum model where
the linear electro-optic effect and
the externally applied microwave field
result in
the interaction between optical cavity modes.
The model assumes
that the number of interacting modes is finite
and
effects of the mode overlapping coefficient
on the strength of the intermode interaction
can be taken into account through dependence of
the coupling coefficient on the mode characteristics.
We show that, under certain conditions,
the model is exactly solvable
and  can be analyzed using the technique of
the Jordan mappings for the $su(2)$ Lie algebra.
Analytical results are applied to
study effects of light modulation on 
the frequency dependence of the photon
counting rate. 
In contrast to the limiting case 
of infinitely large
number of interacting modes, 
when the number of interacting 
modes is finite, the sideband intensities
reveal strongly non-monotonic behavior
supplemented
with asymmetry of the intensity distribution
provided the pumped mode is not central.
We also
analyze different regimes of 
two-modulator transmission and
establish the conditions of validity
of the semiclassical approximation
by applying the methods
of polynomially deformed Lie algebras
for analysis of
the model with quantized microwave field. 
\end{abstract}

\setboolean{displaycopyright}{true}

\begin{document}
\DeclareGraphicsExtensions{.jpeg,.jpg,.eps,.png,.pdf}
\maketitle
\thispagestyle{fancy}

\ifthenelse{\boolean{shortarticle}}{\ifthenelse{\boolean{singlecolumn}}{\abscontentformatted}{\abscontent}}{}

\section{Introduction}
\label{sec:intro}

The quantum information science
being a new rapidly developing branch of 
the modern informatics  
that analyzes how quantum systems
may be used to store, transmit and
process information
heavily relies on  quantum optical information technologies 
where 
units of information are represented by
photons~\cite{Hayashi:bk:2015}. 
Quantum optics is at the heart of 
quantum communication
methods such as quantum cryptography, quantum teleportation, 
and dense 
coding~\cite{Bouwm:bk:2000,Kok:bk:2010,Nielsen:bk:2010,Furusawa:bk:2011,Cariolaro:bk:2015}. 

Quantum photonics and the optical information technology 
provide opportunities for
manipulating the properties of single photons and using them in many
fields~\cite{Yamamoto:pii:2005,Eisaman:rsi:2011,Santori:bk:2010,Migdall:bk:2013,Schiavon:pra:2016}. 
For instance,  
quantum encoding of information underlies
a variety of practical schemes 
developed for quantum cryptography. 
These include
the schemes based on
polarization~\cite{Muller:epl:1993,Muller:nat:1995,Muller:nat:1996}
and  
phase~\cite{Merolla:prl:1999,Merolla:pra:1999} coding,
implementations of quantum cryptography that use
entangled photons~\cite{Ekert:prl:1991,Tittel:prl:2000,Ribordy:pra:2000,Gisin:rmp:2002} 
and
quantum cryptography
protocols based on continuous variables~\cite{Andersen:lpr:2010,Weed:rmp:2012}.

Electro-optical modulators are key devices
for proper operation of 
the above schemes of quantum information processing. 
These devices have also been used in
quantum information methods to monitor and control
different photon quantum 
states~\cite{Merolla:pra:1999,Bloch:ol:2007,Kolchin:prl:2008,Harris:pra:2008,Sensarn:prl:2009,Olislager:pra:2010,Gleim:optexp:2016}.

In solid-state and soft condensed matter physics,
there is a wealth of electro-optic 
effects~\cite{Liu:bk:2005,Yariv:bk:2007,Blin:b:1994,Kiselev:pre:2015,Kiselev:pre:2:2014,Kiselev:ol:2014,Kiselev:pre:2013}  
that underlie the mode of operation of 
a number of optical devices such as modulators, tunable spectral
filters, polarizing converters and optical switches.
The linear electro-optic effect (the Pockels effect)
that occurs
in noncentrosymmetric nonlinear crystals such as
lithium niobate (LiNbO$_3$) crystals
will be of our primary interest.
Though the classical physics
of this effect is well understood~\cite{Liu:bk:2005,Yariv:bk:2007},  
the current and emerging applications of the modulators 
in the field of quantum information
and processing systems  necessitate
developing  quantum approaches to 
frequency and phase modulation~\cite{Tsang:pra:2008,Tsang:pra:2009}.

A consistent quantum theory of phase modulation requires 
the quantum description of the phase.
The problems related to the
quantum phase operator and phase measurements
have an almost century long history dating back to
the original paper by Dirac~\cite{Dirac:prsla:1927}
and
have been the subject of intense studies
(a collection of important papers 
can be found, e.g., in~\cite{Barnett:bk:2007}).
In particular, the quantum theory of
phase and instantaneous frequency along with 
the interferometry methods of measurements 
are described in Refs.~\cite{Tsang:pra:2008,Tsang:pra:2009}.  
In these studies,
quantization of spectrally limited optical fields was performed 
by identifying a slowly varying
envelope of the creation operator and limiting its spectrum to a narrow
band around the carrier frequency.

A quantum scattering theory based black-box approach to
electro-optic modulators is developed in Refs.~\cite{Capmany:josab:2010,Capmany:lpr:2011}. 
In this method, the modulators are regarded as
the scattering devices
producing
a multimode output from a single-mode input.

An alternative approach to phase modulation 
elaborated in early studies~\cite{Yariv:pr:1961,Yariv:bk:2007} 
uses the method of coupled classical modes 
of radiation field
(the classical wave coupling theory
of the electro-optic effect is also discussed in 
Refs~\cite{She:optcomm:2001,Wu:josab:2005}
). 
According to this approach, 
phase modulation of laser radiation
results from
the interaction of cavity eigenmodes caused by 
time periodic modulation of the dielectric
constant of the nonlinear crystal placed in the resonator.
In Ref.~\cite{Kumar:ieeee:2009}, a quantum theory of 
the electro-optic phase modulator is formulated in terms 
of the Hamiltonian describing the intermode interaction 
in the subspace of single photon states.

The common feature of the theoretical considerations
presented
in~\cite{Capmany:josab:2010,Capmany:lpr:2011,Kumar:ieeee:2009}
is that the number of modes
is assumed to be infinitely large
whereas the strength of interaction 
(the coupling coefficient) between the modes
is independent of the mode characteristics.
Though these assumptions greatly simplify theoretical analysis,
they introduce the difficulties related to the unitarity
of the scattering matrix~\cite{Capmany:josab:2010,Capmany:lpr:2011}
and are inapplicable to the case where the modulator
is based on ultra-high quality 
whispering gallery mode 
microresonators
made out of electro-optically active 
materials~\cite{Cohen:ssel:2001,Ilchenko:josab:2003,Savchenkov:ol:2009,Rasoloniaina:scirep:2014}.
  
Such resonators are characterized by
the non-equidistant spectrum of the eigenmodes, 
so that only a small number 
of modes are involved in the interaction
induced by the externally applied microwave field.
The case of three interacting modes 
was theoretically studied
in Refs.~\cite{Ilchenko:josab:2003,Matsko:optexp:2007,Savchenkov:ieee:2010,Tsang:pra:2010}. 
An important result of these studies is that
dependence of the intensity of
sidebands on
the power of the microwave pump
shows the saturation effect
which cannot be explained by
the models where the number
of interacting modes is indefinitely large. 

In this paper our goal is to examine the case
bridging the gap between the above mentioned models
of electro-optic modulators.
For this purpose, we formulate
an exactly solvable model 
in terms of the Hamiltonian describing the parametric process 
where the number of interacting optical modes
is finite and the strength of interaction
varies depending on the mode characteristics
such as the mode number related to the mode frequency.
Owing to algebraic properties of this model,
theoretical analysis can be performed using
the generalized Jordan mapping technique and
the results can be further extended with the help of 
the mathematical methods developed in~\cite{Vadeiko:pra:2003}.

An important point is that, within 
our approach, the modulator is explicitly treated  
as a multiport device (multiport beam splitter)
that may produce and manipulate multiphoton states. 
Such devices are known to be promising for a variety of 
applications~\cite{DellAnno:physrep:2006}. 
In particular, 
the modulator generated multiphoton states 
are used as carriers of information
in
the frequency-coded~\cite{Merolla:prl:1999,Merolla:ol:1999,Bloch:ol:2007} 
and subcarrier multiplexing~\cite{Capmany:pra:2006}
quantum key distribution systems.
 
The paper is organized as follows. 

In Sec.~\ref{sec:model},
we introduce our model and 
show that the mode number dependence
of the coupling coefficient can be reasonably
modeled so as to generate the theory
where the algebraic structure of the multimode operators
is represented by the generators of the $su(2)$ Lie algebra.
In the semiclassical approximation
where the microwave is assumed to be
a classical field, 
analytical expressions for
the evolution operator and the quasienergy spectrum
are obtained in Subsection~\ref{sec:model}.\ref{subsec:operator-evolution}.
In Subsection~\ref{sec:results}.\ref{subsec:photon-counting-rate},
we apply the theoretical results 
to study the effect of light modulation 
on the photon counting rate and present
the results of numerical analysis.
The limiting case where the number of interacting
modes increases indefinitely (the large $S$ limit)
is studied in Subsection~\ref{sec:results}.\ref{subsec:large-number-mode}.
The theory is applied to
analyze different regimes of two-modulator transmission
in Subsection~\ref{sec:results}.\ref{subsec:two-modulator}.
Finally, in Sec.~\ref{sec:concl},
we draw the results together and make some concluding remarks.
Details on the Jordarn mapping technique are relegated to
Appendix~\ref{sec:jordan}.
In Appendix~\ref{sec:quant-mw-field},
we show how the method of polynomially deformed algebras
can be applied to study the quantum model with 
quantized microwave field 
and derive the applicability conditions
for the semiclassical approximation.

\section{Model}
\label{sec:model}

As an electro-optical modulator we consider a nonlinear crystal of
the length $L$ placed between the metal electrodes
parallel to the direction of propagation (the $z$ axis). 
Radio frequency wave
(microwave) excited between the electrodes propagates 
through the crystal along the $z$ axis. 
The microwave mode is characterized
by the wavenumber $k_{\ind{MW}}=2\pi/L$ and the frequency
$\Omega_{\ind{MW}}=k_{\ind{MW}}v_{\ind{MW}}$,
where $v_{\ind{MW}}$ is the phase velocity of the mode.

\begin{figure}[!tbh]
\includegraphics[width=80mm]{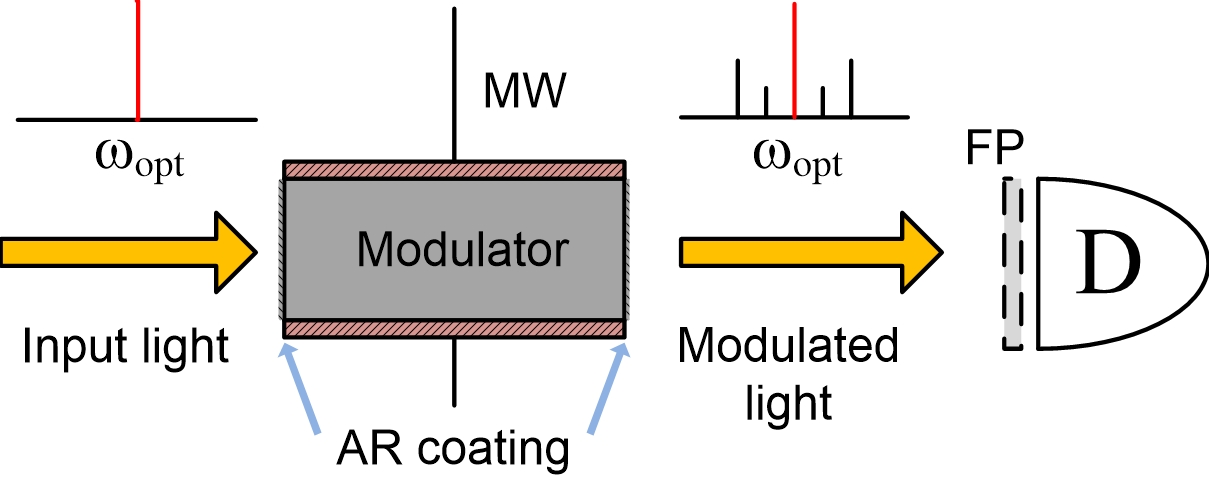}
\caption{%
Modulated light emerging after
the modulator driven by the microwave field (MW)
passes through
a Fabry-Perot filter (FP) and
is collected by a photodetector (D).
Anti-reflective (AR) coating is applied to 
both faces of the electro-optic cavity. 
}
\label{fig:modulator}
\end{figure}

As is illustrated in Fig.~\ref{fig:modulator},
the crystal can be regarded as the reflectionless electro-optic cavity (resonator) 
and we assume that all the traveling optical modes are subject to the periodic
boundary conditions. Then 
the longitudinal wavenumber 
(the $z$ component of the wave vector)
of the modes takes the quantized values:
\begin{align}
  \label{eq:k_n}
  k_m=\frac{2\pi m}{L},
\quad
m\in\mathbb{Z}.
\end{align}
The frequency of the central (carrier) optical mode 
which is typically excited by the laser pulse
is given by
\begin{align}
  \label{eq:omega_opt}
  \omega_{\ind{opt}}=|k_{\ind{opt}}| v_{\ind{opt}}=\Omega
  |m_{\ind{opt}}|,
\quad
\Omega=\frac{2\pi}{L}v_{\ind{opt}},
\end{align}
where 
$k_{\ind{opt}}=\dfrac{2\pi m_{\ind{opt}}}{L}$,
$v_{\ind{opt}}$ is the phase velocity of light in the ambient
dielectric medium and
$m_{\ind{opt}}$ stands for the mode number of 
this operational optical mode. 
Note that the magnitude of the optical mode number $m_{\ind{opt}}$   
is typically of the order $10^4-10^6$ and,
owing to mismatch between the phase velocities
$v_{\ind{MW}}$ and $v_{\ind{opt}}$
the frequency of the microwave mode 
may generally differ from $\Omega$,
$\Omega_{\ind{MW}}=\Omega\, v_{\ind{MW}}/v_{\ind{opt}}\ne\Omega$.

In classical optics, the well-known picture suggests that,
owing to the linear electro-optic effect in the nonlinear crystal,
the externally applied microwave field modulates the phase of the optical wave 
producing
a multimode output observed as the multiple sidebands
that a single optical carrier develops after 
modulation~\cite{Liu:bk:2005,Yariv:bk:2007}. 
The modulation process thus involves interaction of different
optical modes mediated by the microwave field
and the traveling modes appear to be coupled.

The strength of 
the microwave-field-induced intermode coupling
is mainly determined by the two factors:  
(a)~the electro-optic coefficient and  
(b)~the overlapping coefficients
represented by the averages 
of a product of
the spatial distributions
of the modes and the microwave field
over the volume of interaction.
These factors may severely constrain the number of 
efficiently interacting modes. 
For instance,
in
Refs.~\cite{Ilchenko:josab:2003,Bloch:ol:2007,Matsko:optexp:2007,Savchenkov:ieee:2010,Tsang:pra:2010}, 
theoretical considerations 
of electro-optic modulation
are based on the quantum models
with three interacting optical modes.
Our model can be regarded as a generalization of these results.

The electro-optically induced interactions
generally falls within the realms of
the parametric 
processes in nonlinear quantum 
systems
and 
the theoretical technique developed in
early studies on 
this subject~\cite{Yariv:pr:1961,Armstrong:pr:1962,Gordon:pr:1963,Brown:pra:1979,Dodonov:pra:1993}
can be invoked to model them.
Our starting point is the Hamiltonian of the photons
written in the following form:
\begin{align}
&
  \label{eq:Hamilt-0}
  H/\hbar=
\Omega_{\ind{MW}} \hcnj{b} b
+\Omega A_0+\frac{\gamma_0}{f_{\ind{max}}}
\left\{
A_{+} b +
A_{-} \hcnj{b}
\right\},
\\
&
\label{eq:B_0_Apm}
A_0=\sum_{m} m \hcnj{a}_m a_m,
\quad
A_{-}=
\sum_{m}f(m) \hcnj{a}_m a_{m+1},
\quad
A_{+}=\hcnj{A}_{-},
\end{align}
where a dagger will denote Hermitian conjugation,
$\hcnj{b}$ ($b$) is the creation (annihilation)
operator of the photons in the microwave mode,
$\hcnj{a}_m$ ($a_m$) is the creation (annihilation)
operator of the optical photons numbered 
by the mode number $m$,
$\gamma_0$ is the bare intermode coupling constant
(interaction parameter)
and $f(m)/f_{\ind{max}}$ is the normalized function describing the mode
number dependence of 
the intermode interaction strength,
$\displaystyle f_{\ind{max}}=\max_{m}f(m)$.
Note that in our notations the polarization index has been 
dropped. 
It implies that 
all the modes are assumed to be linearly
polarized along a principal axis of the crystal
and we shall restrict our considerations to 
the geometry
where the state of polarization remains intact.

\subsection{Hamiltonian and $su(2)$ algebra}
\label{subsec:hamiltonian}

Formula~(\ref{eq:Hamilt-0})  presents 
the Hamiltonian with the three-boson interaction
written in the rotating wave approximation.
This approximation assumes that
the intermode interaction is dominated by 
the quasiresonant  terms
that
commute with the operator of the linear momentum:
$\displaystyle K/\hbar=k_{\ind{MW}}\hcnj{b} b + \sum_{m} k_m \hcnj{a}_m a_m$
and are slowly varying in the representation of interaction
(they are proportional to
$\exp(\pm i[\Omega-\Omega_{\ind{MW}}]t)$), 
whereas 
the non-resonant terms are of minor importance.
They produce negligibly small effects and hence 
can be disregarded.
 
Another key assumption taken in our model
is that the intensity of the microwave mode
is sufficiently high for its quantum properties to be ignored.
So, it can be described as the classical wavefield.
In this semiclassical approximation,
the creation (annihilation) operator
$b$ ($\hcnj{b}$) is replaced with the c-number
amplitude $B\exp[-i\Omega_{\ind{MW}} t]$
($\cnj{B}\exp[i\Omega_{\ind{MW}} t]\equiv |B|\exp[i(\Omega_{\ind{MW}} t+\phi)]$),
where an asterisk will indicate complex conjugation
and
$\phi$ is the phase of the amplitude $\cnj{B}$,
and the Hamiltonian~(\ref{eq:Hamilt-0})
can be recast into the form:
\begin{align}
&
  \label{eq:Hamilt-s}
  H/\hbar=
\Omega A_0+\frac{\gamma}{f_{\ind{max}}}
\Bigl\{
A_{+}\exp[-i(\Omega_{\ind{MW}} t+\phi)]
\notag
\\
&
+A_{-} \exp[i(\Omega_{\ind{MW}} t+\phi)]
\Bigr\},
\end{align}
where $\gamma=\gamma_0 |B|$.
Applicability of the semiclassical approach
is discussed in Appendix~\ref{sec:quant-mw-field} 
where 
the method of polynomially deformed
algebras developed in Ref.~\cite{Vadeiko:pra:2003}
is used to analyze
the model with quantized microwave mode. 

From the above discussion,
the number of interacting modes
is finite and 
the mode number range for 
these modes can generally
be defined by the inequality
of the form:
\begin{align}
  \label{eq:m-number}
  m_{\ind{min}}<m\le m_{\ind{max}},
\end{align}
where $m_{\ind{min}}$ and $m_{\ind{max}}$
are both the positive integers.
The optical central (operational) 
mode~(\ref{eq:omega_opt})
is in the middle
of the interval~(\ref{eq:m-number})
with the mode number given by
\begin{align}
  \label{eq:m_central}
  m_{\ind{opt}}=\frac{m_{\ind{max}}+m_{\ind{min}}+1}{2}
\end{align}
and we assume that the most efficient intermode coupling
occurs in the vicinity of $m_{\ind{opt}}$,
whereas the strength of interaction
decays to the limit of non-interacting modes at 
the boundaries of the interaction interval~(\ref{eq:m-number}).
Such behavior can be modeled
through the function
\begin{align}
  \label{eq:f_m}
  f(m)=\sqrt{(m-m_{\ind{min}})(m_{\ind{max}}-m)}
\end{align}
describing dependence
of the intermode coupling
on the mode number.

The interacting modes can be conveniently relabeled
by the shifted mode number $\mu$ as follows
\begin{align}
&
  \label{eq:mu}
  m=m_{\ind{opt}}+\mu,
\quad
-S \le \mu \le S,
\quad
S=\frac{m_{\ind{max}}-m_{\ind{min}}-1}{2},
\\
&
  \label{eq:a_mu}
a_{m_{\ind{opt}}+\mu}\equiv a_\mu,
\quad
\hcnj{a}_{m_{\ind{opt}}+\mu}\equiv
\hcnj{a}_\mu,
\end{align}
where $2S+1=m_{\ind{max}}-m_{\ind{min}}=2f_{\ind{max}}$
is the number of interacting modes. 

The operators that enter
the Hamiltonian~(\ref{eq:Hamilt-s})
of our model 
can now be written in the following
form
\begin{align}
&
  \label{eq:A0}
  A_0=m_{\ind{opt}} N + J_0,
\quad
N=\sum_{\mu=-S}^{S}\hcnj{a}_\mu a_\mu,
\quad
J_0=\sum_{\mu=-S}^{S}\mu\hcnj{a}_\mu a_\mu,
\\
&
  \label{eq:J0}
A_{-}\equiv J_{-}=
\sum_{\mu=-S}^{S-1}\sqrt{(S+\mu+1)(S-\mu)}\hcnj{a}_\mu a_{\mu+1},
\\
&
  \label{eq:Jpm}
A_{+}\equiv J_{+}=\hcnj{J}_{-}=
\sum_{\mu=-S}^{S-1}\sqrt{(S+\mu+1)(S-\mu)}\hcnj{a}_{\mu+1} a_{\mu},
\end{align}
where $N$ is the total photon number operator.
The important point is that
the operators $J_{0}$ and $J_{\pm}$
given in Eqs.~(\ref{eq:J0}) and~(\ref{eq:Jpm})
meet 
the well-known commutation relations for
the generators of the $su(2)$ Lie algebra:
\begin{align}
  \label{eq:su2}
  [J_0,J_{\pm}]=\pm J_{\pm},
\quad
[J_{+},J_{-}]=2 J_{0}.
\end{align}
Note that the operator of total photon number
$N$ and the Casimir operator given by
\begin{align}
&
  \label{eq:J2}
  J^2=J_{0}^2+(J_{+} J_{-}+J_{-}J_{+})/2=J_x^2+J_y^2+J_z^2,
\\
&
  \label{eq:Jxyz}
J_z\equiv J_0,\quad
J_{\pm}=J_x\pm i J_y
\end{align}
both commute with the generators of $su(2)$
algebra.

Mathematically, 
the technique of
the generalized Jordan mappings
for bosons~\cite{Biedenharn:bk:1981}
can be applied to derive the relations~(\ref{eq:su2}).
This technique is briefly reviewed in Appendix~\ref{sec:jordan}.

The operators~(\ref{eq:A0})--(\ref{eq:Jpm})
can now be substituted
into Eq.~(\ref{eq:Hamilt-s})
with $f_{\ind{max}}=(2S+1)/2$
to yield the resulting expression for
the time-dependent Hamiltonian of our semiclassical model
\begin{align}
&
  \label{eq:Hamilt-f}
  H(t)/\hbar=\omega_{\ind{opt}} N+
\Omega J_z+\frac{2 \gamma}{2S+1}
\Bigl\{
J_{+}\exp[-i(\Omega_{\ind{MW}} t+\phi)]
\notag
\\
&
+J_{-} \exp[i(\Omega_{\ind{MW}} t+\phi)]
\Bigr\}.
\end{align}

\subsection{Operator of evolution and quasienergy spectrum}
\label{subsec:operator-evolution}

Now we show how the algebraic structure of the model
can be used to evaluate the operator of evolution
and the quasienergy spectrum of the time-periodic 
Hamiltonian~(\ref{eq:Hamilt-f}):
$H(t+T_{\ind{MW}})=H(t)$, where
$T_{\ind{MW}}=2\pi/\Omega_{\ind{MW}}$ is the period
of the microwave mode.

The operator of evolution (propagator) can be found by solving
the initial value problem:
\begin{align}
  \label{eq:op-ev-eq}
  i \hbar \frac{\dd}{\dd t} U(t)=H(t) U(t),
\quad
U(0)=I,
\end{align}
where $I$ is the identity operator.
In the Floquet representation,
the propagator takes the form of a product: 
\begin{align}
  \label{eq:Floq-rep}
  U(t)=U_P(t)\exp(-i Q t/\hbar),
\end{align}
where $U_P(t+T_{\ind{MW}})=U_P(t)$ is the time-periodic operator
and $Q$ is the quasienergy operator.
In our case, equation~(\ref{eq:Floq-rep}) can be
regarded as the rotating wave ansatz
\begin{align}
  \label{eq:RW-ansatz}
  U(t)=U_P(t) U_R(t)
\end{align}
with the unitary operator
\begin{align}
  \label{eq:U_p}
  U_P(t)=\exp[-i (m_{\ind{opt}} N + J_z)\Omega_{\ind{MW}} t]
\end{align}
performing the transformation to the ``rotating coordinate system'' 
and the quasienergy operator is given by
\begin{align}
  \label{eq:Q}
&
  Q/\hbar=
m_{\ind{opt}}\omega  N + \omega J_z
+\frac{2 \gamma}{2S+1}
\Bigl\{
J_{+}\exp(-i\phi)
\notag
\\
&
+J_{-} \exp(i\phi)
\Bigr\},
\quad
\omega=\Omega-\Omega_{\ind{MW}}.
\end{align}

The rotation operator 
\begin{align}
  \label{eq:R}
R(\phi,\beta)=\exp(-i\phi J_z)\exp(-i\beta J_y)
\end{align}
can now be used to
transform the quasienergy operator
into the diagonal form
\begin{align}
&
  \label{eq:Q-d}
Q_d/\hbar =\hcnj{R}(\phi,\beta) Q R(\phi,\beta) 
  = m_{\ind{opt}} \omega  N + \Gamma J_z,
\\
&
\label{eq:Q-param}
\omega+i \frac{4\gamma}{2S+1} = \Gamma\exp[i\beta],
\quad
\Gamma=\sqrt{\omega^2+[4\gamma/(2S+1)]^2},
\end{align}
so that the Fock states 
$\ket{n_{-S},\ldots,n_{\mu},\ldots n_{S}}$
characterized by the mode occupation numbers
are the eigenstates of the quasienergy operator~(\ref{eq:Q-d})
with the quasienergies given by
\begin{align}
  \label{eq:E_quasi}
  E(n,m_z)/\hbar=n\, m_{\ind{opt}} \omega   + m_z \Gamma,
\end{align}
where 
$n=\sum_{\mu=-S}^{S}n_{\mu}$ is the total number of photons and 
$m_z=\sum_{\mu=-S}^{S}\mu n_{\mu}$ is the azimuthal quantum number,
$-nS \le m_z\le nS$.
Note that a set of
the Fock states characterized by the quantum numbers
$n$ and $m_z$ forms a vector space $\mathcal{F}_{n,\,m_z}$ which
can further be divided into subspaces $\mathcal{F}_{n,\,m_z}^{j}$ classified by the eigenvalues
of the Casimir operator~(\ref{eq:J2}):
$J^2\ket{n,j,m_z,\kappa}=j(j+1)\ket{n,j,m_z,\zeta}$,
where $|m_z|\le j\le nS$ is the angular momentum
number and $\zeta$ is the integer enumerating
the basis eigenstates of $\mathcal{F}_{n,\,m_z}^{j}$,
$\kappa\in\{1,\ldots, \dim \mathcal{F}_{n,\,m_z}^{j}\}$.
In other words, the space $\mathcal{F}_{n,\,m_z}$ can generally be decomposed
into the direct sum of subspaces $\mathcal{F}_{n,\,m_z}^{j}$:
$\mathcal{F}_{n,\,m_z}=\underset{j}{\oplus}{\mathcal{F}_{n,\,m_z}^{j}}$.
For example, at $S=1$, it can be shown that
$0\le j= n-2k\le n$ with $k\in\{0,1,\ldots\}$ and all the subspaces $\mathcal{F}_{n,\,m_z}^{j}$
are one-dimensional, $\dim \mathcal{F}_{n,\,m_z}^{j}=1$.
Given the total photon number $n$,
the dimension of the space $\mathcal{F}_n=\underset{m_z}{\oplus}{\mathcal{F}_{n,\,m_z}}$
is known to be $\dim \mathcal{F}_n=(n+2S)![(2S)! n!]^{-1}$
and, at $S>1$, the subspaces $\mathcal{F}_{n,\,m_z}^{j}$
are not necessarily one-dimensional. 

Equations~(\ref{eq:U_p})--~(\ref{eq:Q-d})
can now be substituted
into the Floquet representation~(\ref{eq:Floq-rep})
to yield the operator of evolution in the final form:
\begin{align}
  \label{eq:U-result}
  U(t)=\e^{-i\Omega_{\ind{MW}}t J_z}R(\phi,\beta)\e^{-i\Gamma t J_z}\hcnj{R}(\phi,\beta)\e^{-i\omega_{\ind{opt}}t N}. 
\end{align}
Using this formula in combination with
the identity for the rotation
of the annihilation operator~(\ref{eq:a-R})
derived in Appendix~\ref{sec:jordan},
we can describe the temporal evolution
of the photon annihilation operator
$a_{\mu}$ as follows
\begin{align}
&
  \label{eq:a-U}
  a_{\mu}(T)=\hcnj{U}(T) a_{\mu} U(T)
=\sum_{\nu=-S}^{S}
M_{\mu\nu} a_{\nu},
\\
& 
  \label{eq:a-M}
M_{\mu\nu}=\e^{-i(\omega_{\ind{opt}}+\mu\Omega_{\ind{MW}}) T}
\e^{-i(\mu-\nu)\phi}
U_{\mu\nu}^{S}(T),
\\
&
\label{eq:U-s}
U_{\mu\nu}^{S}(T)=
\sum_{\mu'=-S}^{S}
d_{\mu\mu'}^{S}(\beta)
d_{\nu\mu'}^{S}(\beta)\e^{-i\mu'\Gamma T}=(-1)^{\nu}\e^{-i(\mu+\nu)\tilde\alpha}d_{\mu\nu}^{S}(\tilde\beta),
\end{align}
where 
$T$ is the duration of intermode interaction
which is the time an optical wavefield takes 
to propagate through the
electro-optic modulator
and
the expression for $U_{\mu\nu}^{S}(T)$ 
is simplified by using
the addition formula for the Wigner
$D$~functions~\cite{Varshalovich:bk:1988},
$D_{\mu\nu}^{S}(\alpha,\beta,\gamma)=\exp[-i\mu\alpha]d_{\mu\nu}^{S}(\beta)\exp[-i\nu\gamma]$, 
with the angles $\tilde\alpha$ and $\tilde{\beta}$ defined through the
following relations:
\begin{subequations}
\label{eq:talp-tbeta}
\begin{align}
&
  \label{eq:talp}
  2\tilde\alpha=\pi+\arg\{\sin^2\beta+(1+\cos^2\beta)\cos(\Gamma T)
+ 2i
  \cos\beta\sin(\Gamma T)\},
\\
&
  \label{eq:c-tbeta}
\cos\tilde\beta=\cos^2\beta+\sin^2\beta\cos(\Gamma T),
\\
&
  \label{eq:s-tbeta}
\sin\tilde\beta=\sin\beta\bigr[
\cos\tilde\alpha \cos\beta (1-\cos(\Gamma T))
-\sin\tilde\alpha\,\sin(\Gamma T)
\bigr].
\end{align}
\end{subequations}
Details on derivation of these relations are relegated
to Appendix~\ref{sec:jordan}.

\section{Results}
\label{sec:results}

The model described
in the previous section
represents the electro-optic modulator
as a multiport device that
may generally be used to 
manipulate multimode photonic states.
Quantum dynamics of such states 
involves different frequency modes 
and lie at the heart of applications 
based on frequency encoding~\cite{Merolla:pra:1999,Bloch:ol:2007,Gleim:optexp:2016}
and quantum effects dealing with
the frequency entangled photons~\cite{Olislager:pra:2010,Olislager:njp:2012}.
For these applications and effects
an  important fisrt step is to 
study the effect of light modulation 
on the photon counting rate.

For this purpose, in this section, 
we use the analytical results
of Sec.~\ref{sec:model}
to evaluate the counting rate as
the one-electron photodetection probability
per unit time.
We find that it
can be written in the factorized form with 
the modulation form-factor expressed in terms of
the matrix~(\ref{eq:U-s})
and present a number of
numerical results for this form-factor.
We also discuss what happen in
the limiting case where the number of interacting
modes increases indefinitely and $S\to\infty$
(the large $S$ limit) and
apply the theory to the problem of 
two-modulator transmission.

\subsection{Photon counting rate}
\label{subsec:photon-counting-rate}

The cavity modes of the modulator
are excited by the photons with the carrier frequency 
$\omega_{\ind{opt}}$
that propagate through the electro-optic device. 
Owing to the electro-optic effect,
the traveling microwave field inside the cavity
gives rise to the intermode interaction.
For the modes initially
prepared in the state described by the
density matrix of the radiation field $\rho_F(0)$,
at the instant of time $T$,
the density matrix is given by
\begin{align}
  \label{eq:rho_T}
  \rho_F(T)=U(T)\rho_F(0)\hcnj{U}(T),
\end{align}
where $U(t)$ is the operator of evolution~(\ref{eq:U-result})
(the losses are neglected).
Note that, similar to Eq.~(\ref{eq:a-U}), 
$T$ is the duration of interaction
(the time it takes for a light wave to travel through 
the region where electro-optic modulation occurs)
and
Eq.~(\ref{eq:rho_T}) assumes 
the lossless dynamics of the density matrix $\rho_F$. 

An important characteristics of the radiation field 
is the averaged photon number
of the mode with the mode number $\mu$:
\begin{align}
  \label{eq:n-avr}
\avr{N_{\mu}}(T)=\Tr_F\{\hcnj{a}_{\mu}a_{\mu}\rho_F(T)\}=
\Tr_p\{\hcnj{a}_{\mu}(T)a_{\mu}(T)\rho_F(0)\}.
\end{align}
We can now use the relation~(\ref{eq:a-U})
to derive the explicit expression for the average~(\ref{eq:n-avr}):
\begin{align}
  \label{eq:n-avr-gen}
  \avr{N_{\mu}}(T)=
\sum_{\nu',\nu=-S}^{S}
U_{\mu\nu'}^{S\,*}(T)
U_{\mu\nu}^{S}(T)
\e^{i(\nu-\nu')\phi}
\avr{\hcnj{a}_{\nu'}a_{\nu}}(0)
\end{align}
where
$\avr{\hcnj{a}_{\nu'}a_{\nu}}(0)=\Tr_F\{\hcnj{a}_{\nu'}a_{\nu}\rho_F(0)\}$.
For  the special case of single mode pumping where
the optical mode with the mode number $\nu$ is the only mode
initially excited in the resonator
and $\avr{\hcnj{a}_{\nu'}a_{\nu}}(0)=\delta_{\nu\nu'} \avr{N_{\nu}}(0)$, 
we have
\begin{align}
  \label{eq:n-avr-single}
  \avr{N_{\mu}}(T)\equiv \avr{N_{\mu}}=
|U_{\mu\nu}^{S}(T)|^2
  \avr{N_\nu}(0).
\end{align} 
Typically, 
this is the central mode which is excited
and $\nu=0$.

We consider an experimental setup 
depicted in Fig.~\ref{fig:modulator}.
In such a setup,
the output of the electro-optic modulator 
is connected to a Fabry-Perot filter via the optical
fiber channel. Then the light wave passed through the filter
is collected by a photodetector with sufficiently wide bandwidth.  

The wavefield at the exit of the modulator
is characterized by the density matrix
$\rho_{F}(T)$ given in Eq.~(\ref{eq:rho_T}).
An important point is that
the modes of 
the output light field should be matched 
to the modes of the optical fiber. 
 In what follows we shall assume that 
they are perfectly correlated
(see, e.g., Chapter~1.4 in the book~\cite{Carmichael:bk:1993}). 
So, in the interaction picture,
the electric field operator,
$\vc{E}(\vc{r},t)$,
of light normally incident on the filter
can be decomposed into its positive and negative
frequency parts,
 $\vc{E}_{+}(\vc{r},t)$ and $\vc{E}_{-}(\vc{r},t)$,
as follows 
\begin{align}
  \label{eq:E-operator}
&
  \vc{E}(\vc{r},t)=\vc{E}_{+}(\vc{r},t)+\vc{E}_{-}(\vc{r},t),
\notag
\\
&
\vc{E}_{+}(\vc{r},t)=\hcnj{\vc{E}}_{-}(\vc{r},t)=
\sum_{\mu=-S}^{S}\vc{E}_{\mu}^{(+)}(\vc{r}) a_{\mu}(t),
\end{align}
where $a_{\mu}(t)=a_{\mu}(T)\exp[-i\omega_{\mu} (t-T)]$,
$a_{\mu}(T)$ is given in Eq.~(\ref{eq:a-U}),
$\omega_{\mu}=\omega_{\ind{opt}}+\mu\Omega$
is the mode frequency and $\vc{E}_{\mu}^{(+)}(\vc{r})$ is
the complex valued vector amplitude that
generally depends on 
a number of characteristics 
of the mode such as 
the wavevector, the frequency and 
the state of polarization.
Throughout this paper we have 
used notations where
the index describing
the state of polarization of the
modes is suppressed
by assuming that
all the modes are linearly polarized
along the unit vector $\uvc{e}$,
so that $\vc{E}_{\mu}^{(+)}(\vc{r})={E}_{\mu}^{(+)}(\vc{r})\uvc{e}$.

For the light transmitted through the filter,
$\vc{E}_{\ind{f}}=\vc{E}_{+}^{(\ind{f})}+\vc{E}_{-}^{(\ind{f})}$,
we have
\begin{align}
  \label{eq:E-trans}
  \vc{E}_{+}^{(\ind{f})}(\vc{r},t)=
\sum_{\mu=-S}^{S}\tilde{\vc{E}}_{\mu}^{(+)}(\vc{r})
a_{\mu}(t),
\:
\tilde{\vc{E}}_{\mu}^{(+)}(\vc{r})=
\mvc{T}_{\ind{f}}(\omega_f,\omega_{\mu})\vc{E}_{\mu}^{(+)}(\vc{r}),
\end{align}
where $\mvc{T}_{\ind{f}}(\omega_f,\omega_{\mu})$
is the transmission matrix of the filter
and $\omega_f$ is the filter frequency.
The filter is also characterized by the bandwidth
$\Delta\omega_f$, so that
the filter transmission is negligibly small
provided that
$|\omega_f-\omega_{\mu}|>\Delta\omega_f$.

We now briefly discuss 
the process of photoelectric detection of
the light transmitted through the filter
based on the model of an idealized 
photodetector described in the monograph~\cite{Mandl:bk:1995}
(Chapter~14).
Our task is to compute
the photon counting rate as
the one-electron photodetection probability
per unit time.
For this purpose, we take the assumption of
a narrowband optically isotropic
filter with $\Delta\omega_f<\Omega$
and $\mvc{T}_{\ind{f}}(\omega_f,\omega_{\mu})=
T_{\ind{f}}(\omega_f,\omega_{\mu})\mvc{I}_3$,
where $\mvc{I}_3$ is the $3\times 3$ identity matrix.
So, the result can now be obtained by using the quasi-monochromatic
approximation (see Chapter~14.2.2 in the book~\cite{Mandl:bk:1995}).
For an atom located at $\vc{r}=\vc{r}_0$, 
the energy of interaction between the atom and the radiation
field in the dipole approximation is  
\begin{align}
  \label{eq:V}
  V=-\sca{\vc{d}}{\vc{E}_{\ind{f}}(\vc{r}_0)},
\end{align}
where $\vc{d}$ is the operator of the electric dipole moment.

We can now closely follow the line of reasoning
presented in Ref.~\cite{Mandl:bk:1995}
and obtain the one-electron detection probability rate
\begin{align}
  \label{eq:P-1}
  p(\omega_f)=\sum_{\mu=-S}^{S}|T_{\ind{f}}(\omega_f,\omega_{\mu})|^2
\avr{N_{\mu}} K(\omega_{\mu})
\end{align}
expressed in terms of the frequency
response function of the photodetector
\begin{align}
\label{eq:K-omega}
K(\omega_{\mu})=&
H(\omega_{\mu}-\omega_g)
\int
\sigma(\omega_{\mu}-\omega_g,\kappa) g(\omega_{\mu}-\omega_g,\kappa)
\notag
\\
&
\times |\bra{\omega_{\mu}-\omega_g,\kappa}\sca{\vc{d}}{\uvc{e}}\ket{G}{E}_{\mu}^{(+)}(\vc{r}_0)|^2
\dd\kappa,  
\end{align}
where $H(x)$ is the Heaviside unit step function,
$\ket{G}$ is the ground (bounded) state of the atom
with the negative energy equal to $-\hbar\omega_g$ 
(it is the eigenstate of the Hamiltonian of the atomic system
$H_A$: $H_A\ket{G}=-\hbar\omega_g\ket{G}$),
$\ket{\omega_e,\kappa}$ is the excited 
free electron (unbound) state 
characterized by the positive energy $\hbar\omega_e$
($H_A\ket{\omega_e,\kappa}=\hbar\omega_e\ket{\omega_e,\kappa}$)
and possibly by other variables represented by $\kappa$;
$\sigma(\omega_e,\kappa)$ is the density of the excited states
and $g(\omega_e,\kappa)$ is the probability for
the electron in the state $\ket{\omega_e,\kappa}$
to be collected and registered by the detector.

For the broadband detector with
$K(\omega_{\mu})\approx K(\omega_{\ind{opt}})$,
the expression for the counting rate~(\ref{eq:P-1})
can be further simplified giving the result in 
the factorized form:
\begin{align}
&
  \label{eq:P-2}
  p(\omega_f)\approx  p_0(\omega_{\ind{opt}}) p_{\ind{mod}}(\omega_f,T),
\notag
\\
&
p_0(\omega_{\ind{opt}})=K(\omega_{\ind{opt}}) \avr{N_\nu}(0),
\\
&
\label{eq:p0-rel}
p_{\ind{mod}}(\omega_f,T)=
\sum_{\mu=-S}^{S}|T_{\ind{f}}(\omega_f,\omega_{\mu})
U_{\mu\nu}^{S}(T)|^2,
\notag
\\
&
|U_{\mu\nu}^{S}(T)|^2=|d_{\mu\nu}^{S}(\tilde\beta)|^2,
\end{align}
where we have used formulas~(\ref{eq:n-avr-single})
and~(\ref{eq:U-s})
for the averaged photon number $\avr{N_{\mu}}$
and $U_{\mu\nu}^{S}(T)$, respectively. 
From Eqs.~(\ref{eq:P-2}) and~(\ref{eq:p0-rel}),
it is clear that the photon count form-factor
$p_{\ind{mod}}(\omega_f,T)$ accounts for the combined effect of
the modulator and the filter,
whereas the factor $p_0(\omega_{\ind{opt}})$ 
gives the counting rate without filtering and modulation. 
The form-factor $p_{\ind{mod}}(\omega_f,T)$
thus might be called the light modulation form-factor
of the photon count rate.  

\begin{figure}[!tbh]
\includegraphics[width=80mm]{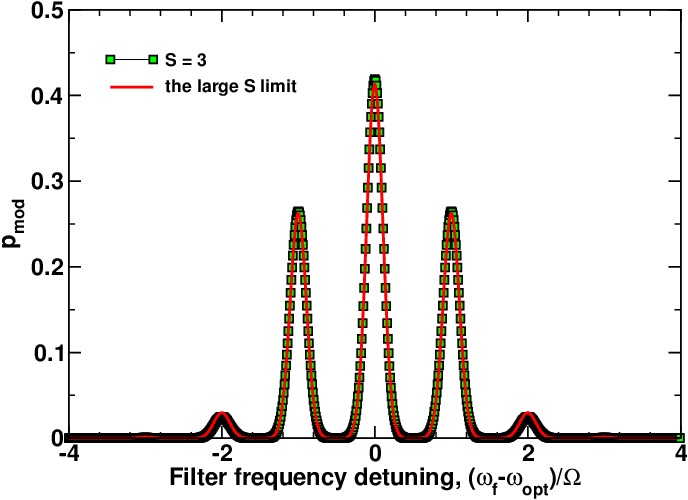}
\caption{%
(Color online)
 The photon counting rate form-factor $p_{\ind{mod}}$ as a function of 
the filter frequency detuning
computed from Eq.~(\ref{eq:p0-rel})
at the intermode coupling parameter
$\gamma/\Omega=0.1$
(the regime of weak intermode coupling).
The other parameters are:
$T_{\ind{max}}^{(f)}=1$ 
is the maximal transmittance of the filter;
$\sigma_f/\Omega=0.15$
is the bandwidth of the filter;
$T= 2\pi/\Omega$ is the time of intermode interaction
and
$\omega/\Omega=(\Omega-\Omega_{\ind{MW}})/\Omega=0.01$. 
}
\label{fig:count-f-1}
\end{figure}

\begin{figure}[!tbh]
\includegraphics[width=80mm]{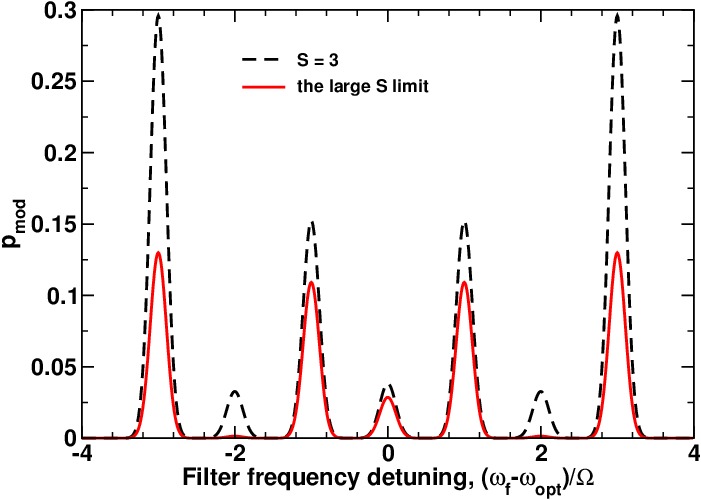}
\caption{%
(Color online)
 The photon counting rate form-factor $p_{\ind{mod}}$ as a function of 
the filter frequency detuning
at the intermode coupling parameter
$\gamma/\Omega=0.4$
(the regime of intermediate intermode coupling).
Other parameters are listed in the caption
of Fig.~\ref{fig:count-f-1}. 
}
\label{fig:count-f-2}
\end{figure}

\begin{figure}[!tbh]
\includegraphics[width=80mm]{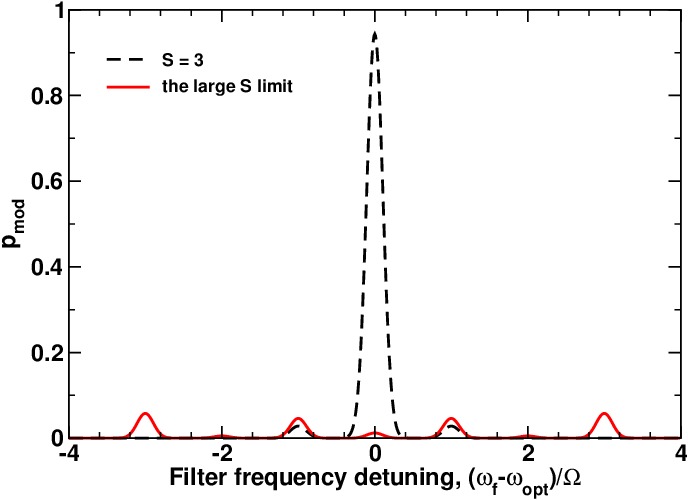}
\caption{%
(Color online)
 The photon counting rate form-factor $p_{\ind{mod}}$ as a function of 
the filter frequency detuning
at the intermode coupling parameter
$\gamma/\Omega=0.9$
(the regime of strong intermode coupling).
Other parameters are listed in the caption
of Fig.~\ref{fig:count-f-1}. 
}
\label{fig:count-f-3}
\end{figure}

\begin{figure}[!tbh]
\includegraphics[width=80mm]{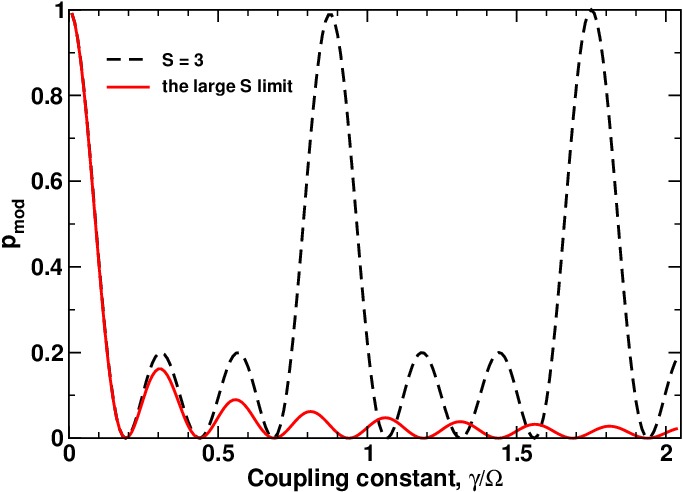}
\caption{%
(Color online)
 The photon counting rate form-factor $p_{\ind{mod}}$ as a function of 
the coupling parameter $\gamma/\Omega$
at $\omega_f=\omega_{\ind{opt}}$.
Other parameters are listed in the caption
of Fig.~\ref{fig:count-f-1}. 
}
\label{fig:count-g-1}
\end{figure}

\begin{figure}[!tbh]
\includegraphics[width=80mm]{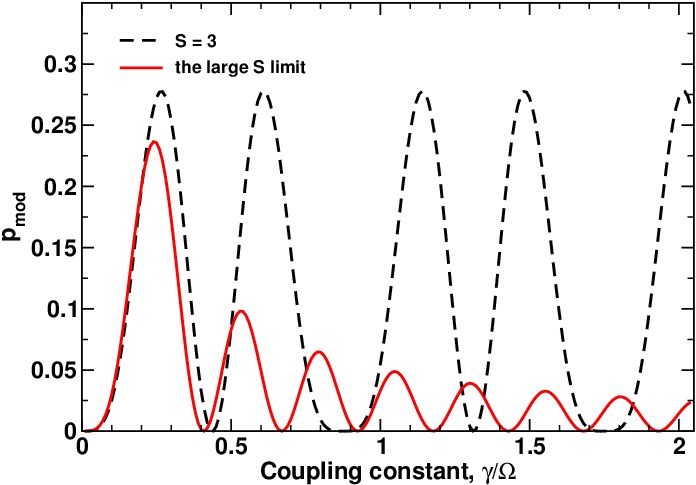}
\caption{%
(Color online)
 The photon counting rate form-factor $p_{\ind{mod}}$ as a function of 
the coupling parameter $\gamma/\Omega$
at $\omega_f=\omega_2=\omega_{\ind{opt}}+2\Omega$.
Other parameters are listed in the caption
of Fig.~\ref{fig:count-f-1}. 
}
\label{fig:count-g-2}
\end{figure}

In our calculations, the frequency dependence of 
the filter transmittance
is modeled by the Gaussian shaped curve
\begin{align}
  \label{eq:T-f}
  |T_{\ind{f}}(\omega_f,\omega)|^2=
T_{\ind{max}}^{(f)}
\exp[-(\omega_f-\omega)^2/\sigma_f^2],
\end{align}
where $\sqrt{\ln 2}\, \sigma_f$ is 
the Gaussian half width at half maximum
that determines the bandwidth of the filter $\Delta\omega_f=\sigma_f$
and 
$T_{\ind{max}}^{(f)}$ is the maximal transmittance of 
the filter at the peak $\omega_f=\omega$.
Figures~\ref{fig:count-f-1}--\ref{fig:count-g-2}
show the photon count light modulation form-factor
$p_{\ind{mod}}(\omega_f,T)$
computed from Eq.~(\ref{eq:p0-rel}) either as a function of the dimensionless filter frequency
detuning $(\omega_{f}-\omega_{\ind{opt}})/\Omega$
(Figs.~\ref{fig:count-f-1}--\ref{fig:count-f-3})
or in relation to the intermode coupling parameter
$\gamma/\Omega$
(Figs.~\ref{fig:count-g-1} and~\ref{fig:count-g-2}).
In these figures, 
the mode initially excited in the electro-optic cavity
is central with $\nu=0$ and
the parameters are:
$T_{\ind{max}}^{(f)}=1$, $\sigma_f/\Omega=0.15$,
$\omega/\Omega=0.01$ and $T= 2\pi/\Omega$.

\subsection{
Regime of large number of interacting modes:
the large $S$ limit}
\label{subsec:large-number-mode}

In our model, 
the operator of evolution~(\ref{eq:U-result})
describing
the effect of electro-optically induced light modulation
is represented by the matrix $U_{\mu\nu}^{S}(T)$
given by Eq.~(\ref{eq:U-s}).
The latter is the $(2S+1)\times (2S+1)$ matrix,
where $2S+1$ is 
the number of interacting modes. 
In this section, we discuss the limiting case
where the number of interacting modes
is large and $S\to\infty$.

From Eq.~(\ref{eq:U-s}),
the elements of the matrix $U_{\mu\nu}^{S}(T)$
are determined by the Wigner $D$~functions
$D_{\mu\nu}^{S}(\tilde\alpha,\tilde\beta,\tilde\alpha+\pi)$
with relations for 
the angles $\tilde\alpha$ and $\tilde\beta$
given by Eq.~(\ref{eq:talp-tbeta}). 
Equation~(\ref{eq:Q-param})
can be used in combination 
with the relations~(\ref{eq:talp-tbeta})
to derive the large $S$ asymptotics
for the angles
\begin{subequations}
\label{eq:angles-asymp}
\begin{align}
&
  \label{eq:beta-asymp}
  \sin\beta\sim \frac{2\gamma}{|\omega|} S^{-1},
\\
&
\label{eq:talp-asymp}
\tilde\alpha\xrightarrow[S\to\infty]{} \frac{\pi+\omega T}{2},
\quad \omega=\Omega-\Omega_{\ind{MW}},
\\
&
\label{eq:tbet-asymp}
\tilde\beta\sim -\frac{g}{S},
\quad
g=\frac{4\gamma}{|\omega|}\sin(|\omega| T/2).
\end{align}
\end{subequations}

Our next step
starts with 
the well-known expression for
the Wigner $d$-functions~\cite{Biedenharn:bk:1981}
\begin{align}
&
  \label{eq:d-func}
  d_{\mu\nu}^{S}(\tilde\beta)=\sqrt{\frac{(S+\nu)!(S-\nu)!}{(S+\mu)!(S-\mu)!}}
\sin^{\nu-\mu}\left(\frac{\tilde\beta}{2}\right)
\notag
\\
&
\times
\cos^{\nu+\mu}\left(\frac{\tilde\beta}{2}\right)
P_{S-\nu}^{(\nu-\mu,\nu+\mu)}(\cos\tilde\beta),
\end{align}
where $P_n^{(\alpha,\beta)}(x)$ denotes the Jacobi
polynomial~\cite{Abr},
and uses asymptotics of the Jacobi polynomials
given by the Mehler-Heine formula~\cite{Szego:bk:1975}:
\begin{align}
&
  \label{eq:Mehler}
\lim_{n\to \infty} n^{-\alpha} P_n^{(\alpha,\beta)}(\cos(z/n))=
  \lim_{n\to \infty} n^{-\alpha} P_n^{(\alpha,\beta)}(1-z^2n^{-2}/2)
\notag
\\
&
=
\left[\frac{z}{2}\right]^{-\alpha} J_\alpha(z),
\end{align}
where $J_\alpha(z)$ is the Bessel function
of the first kind~\cite{Abr}
(outside this subsection symbols 
$J_\mu$ without arguments denote the generators of $su(2)$).

From Eqs.~(\ref{eq:angles-asymp})--~(\ref{eq:Mehler}),
it is rather straightforward to find that 
the asymptotic behavior of the Wigner $d$~functions
and the matrix $U_{\mu\nu}^{S}$ is given by
\begin{align}
&
  \label{eq:d-func-asymp}
  d_{\mu\nu}^{S}(\tilde\beta)\xrightarrow[S\to\infty]{}J_{\mu-\nu}(g),
\\
&
  \label{eq:U-asymp}
  U_{\mu\nu}^{S}(T)\xrightarrow[S\to\infty]{}(-i)^{\mu-\nu}
  \e^{-i(\mu-\nu)\omega T/2} J_{\mu-\nu}(g) \e^{-i\nu\omega T}.
\end{align}

Now we apply the asymptotic relation~(\ref{eq:U-asymp})
to describe, in the large $S$ limit, 
the effect of electro-optic modulation on
temporal evolution of light after passing through
the modulator at $t>T$. 
From Eq.~(\ref{eq:E-operator}),
the averaged positive frequency part of the electric field
can be written as follows
\begin{align}
  \label{eq:E-plus-ini}
\avr{\vc{E}_{+}(\vc{r},t)}=
\sum_{\mu=-S}^{S}\vc{E}_{\mu}^{(+)}(\vc{r}) \e^{-i\omega_{\mu}(t-T)}
\avr{a_{\mu}(T)}_0,
\end{align}
where 
$\omega_{\mu}=\omega_{\ind{opt}}+\mu\Omega$ and
$\avr{a_{\mu}(T)}_0\equiv \Tr_F\{a_{\mu}(T)\rho_F(0)\}$.

\begin{align}
&
  \label{eq:E-plus-asymp}
  \avr{\vc{E}_{+}(\vc{r},t)}
\xrightarrow[S\to\infty]{}
\sum_{\mu,\nu}
\vc{E}_{\mu}^{(+)}(\vc{r}) 
\e^{-i(\mu-\nu)(\Omega t-\psi)}
J_{\mu-\nu}(g)\e^{-i\omega_{\nu} t}
\avr{a_{\nu}}_0
\notag
\\
&
\approx 
\sum_{\mu,\nu}
\e^{-i(\mu-\nu)(\Omega t-\psi)}
J_{\mu-\nu}(g)
\vc{E}_{\nu}^{(+)}(\vc{r}) 
\e^{-i\omega_{\nu} t}
\avr{a_{\nu}}_0
\notag
\\
&
=
\e^{-i g\cos (\Omega t - \psi)}
\avr{\vc{E}_{+}(\vc{r},t)}_0,
\end{align}
where
$\psi=\omega T/2-\phi$
and $\avr{\vc{E}_{+}(\vc{r},t)}_0=
\sum_{\nu}\vc{E}_{\nu}^{(+)}(\vc{r}) \e^{-i\omega_{\nu}t}
\avr{a_{\nu}}_0$
is the average of the radiation field
propagating in the free space without light modulation.
The result~(\ref{eq:E-plus-asymp}) is obtained 
with the help of the equality~\cite{Abr}
\begin{align}
  \label{eq:J-rel}
\exp[-ig\cos(\Omega t)]=
  \sum_{\mu=-\infty}^{\infty}(-i)^{\mu}\e^{-i\mu \Omega t} J_{\mu}(g)
\end{align}
by assuming that
the modes are linearly polarized
$\vc{E}_{\mu}^{(+)} =E_{\mu}^{(+)} \uvc{e}$
and the approximation $E_{\mu}^{(+)}\approx E_{\nu}^{(+)}$
may break only in the region where $|\mu-\nu|$ is sufficiently large
for $|J_{\mu-\nu}(g)|$ to be negligibly small.

The phase factor 
$\exp[-i g\cos (\Omega t - \psi)]$
on the right hand side of Eq.~(\ref{eq:E-plus-asymp})
implies that the wave after the modulator
becomes phase modulated and $g$
plays the role of the phase modulation index
(the modulation depth).
This is the well known result of the simple
classical model~\cite{Yariv:bk:2007}
which in our model is recovered
in the large $S$ limit.

In Figs.~\ref{fig:count-f-1}--\ref{fig:count-f-3},
the filter frequency dependence
of the photon count modulation form-factor
computed in the large $S$ limit
is compared with $p_{\ind{mod}}$ 
evaluated at $S=3$
(the number of interacting modes equals $7$).
As is illustrated in Fig.~\ref{fig:count-f-1},
in the case of weak intermode interaction where
the coupling constant is small,
the differences between the curves are negligible.
Referring to  Figs.~\ref{fig:count-f-2} and~\ref{fig:count-f-3},
the latter is no longer the case
in the regimes of intermediate and strong coupling.

The effect of electro-optically induced intermode interaction
can be clearly seen in Figs.~\ref{fig:count-g-1}
and~\ref{fig:count-g-2} where
the form-factor of the mode 
with the frequency $\omega_{\mu}$ 
selected by the filter at $\omega_f=\omega_{\mu}$
is plotted as a function of the coupling constant 
$\gamma$.
For the central mode with $\mu=0$,
the results are presented in Fig.~\ref{fig:count-g-1}.

Clearly, in the large $S$ limit, 
the coupling constant dependence of $p_{\ind{mod}}$
shown in  Fig.~\ref{fig:count-g-1}
demonstrates that the initially pumped mode becomes 
depleted as the strength of interaction increases,
so that the photons spread over the (infinitely) large
number of modes.
By contrast, the model with $S=3$ predicts
qualitatively different behavior of the form-factor
characterized by oscillations with $p_{\ind{mod}}$ being close
to a periodic function of $\gamma$.

Mathematically, the oscillating behavior of $p_{\ind{mod}}$
is determined by the elements of the matrix~(\ref{eq:U-s})
where $|U_{\mu\nu}^{S}|=|d_{\mu\nu}^{S}(\tilde\beta)|$
is a $2\pi$ periodic even function of $\tilde\beta$
with $|d_{\mu\nu}^{S}(0)|=\delta_{\mu\nu}$
and
$|d_{\mu\nu}^{S}(\pi)|=\delta_{\mu,-\nu}$.
In addition, from Eq.~(\ref{eq:talp-tbeta}),
it can be shown that, given the angle $\beta$,
the angle $\tilde\beta$
can be regarded as a function of  $\Gamma T$ and
$|\tilde\beta(\beta,\Gamma T)|=|\tilde\beta(\beta,2\pi\pm\Gamma T)|$.
This implies that $|U_{\mu\nu}^{S}|=|d_{\mu\nu}^{S}(\tilde\beta)|$
is a periodic function of $\Gamma T$.

From Eq.~(\ref{eq:Q-param}),
it can be inferred that, in general,
the parameters $\beta$ and $\Gamma$
both depend on the coupling constant $\gamma$.
At $\gamma\gg |\omega|$,
$\beta\approx\pi/2$ and $\Gamma$ is linearly proportional
to $\gamma$.
So, at sufficiently strong coupling
$|U_{\mu\nu}^{S}|$ (and thus $p_{\ind{mod}}$)
will be a periodic function of $\gamma$.

In particular, when the phase velocities of microwave and optical fields
are matched and $\omega=\Omega-\Omega_{\ind{MW}}=0$,
we deal with the resonance case 
where $\beta=\pi/2$, $\tilde\alpha=-\pi/2$
and $\tilde\beta=\Gamma T=4\gamma T/(2S+1)$
(see Eq.~(\ref{eq:sin-0}) in Appendix~\ref{sec:jordan}).
In the large $S$ limit, it is not difficult to show that
$U_{\mu\nu}^{S}\to (-i)^{\mu-\nu} J_{\mu-\nu}(2\gamma T)$
and we obtain the result in the form of Eq.~(\ref{eq:E-plus-asymp})
with $\psi=-\phi$ and $g=2\gamma T$.
For finite number of modes,
the $\gamma$ dependence of the photon counting rate
is dictated by the coupling dependent factor 
$|d_{\mu\nu}^{S}|^2$.
This factor is an even $2\pi$ periodic function
of the coupling parameter $4\gamma T/(2S+1)$.
In contrast, oscillations of the factor 
$|J_{\mu-\nu}(2\gamma T)|^2$ rapidly
decay in magnitude as $\gamma$ increases.
Figure~\ref{fig:count-g-2} illustrates that 
similar effects occur when
the detuning $\omega$ is small
and $\mu=2$. 

Figures~\ref{fig:count-f-1}--\ref{fig:count-g-2}
present the results obtained by assuming
that the mode excited in the cavity is central
with $\nu=0$.
In this case the model and its large $S$ limit 
both predict that 
$|U_{\mu, 0}^S|^2=|U_{-\mu, 0}^S|^2$
and contributions to the photon counting rate
coming from
symmetrically arranged sideband modes,
$\mu$ and $-\mu$,
are equal. 
This symmetry is evident from
the curves shown in Figs.~\ref{fig:count-f-1}--\ref{fig:count-f-3}.

\begin{figure*}[!tbh]
\centering
\subfloat[]{
  \resizebox{70mm}{!}{\includegraphics*{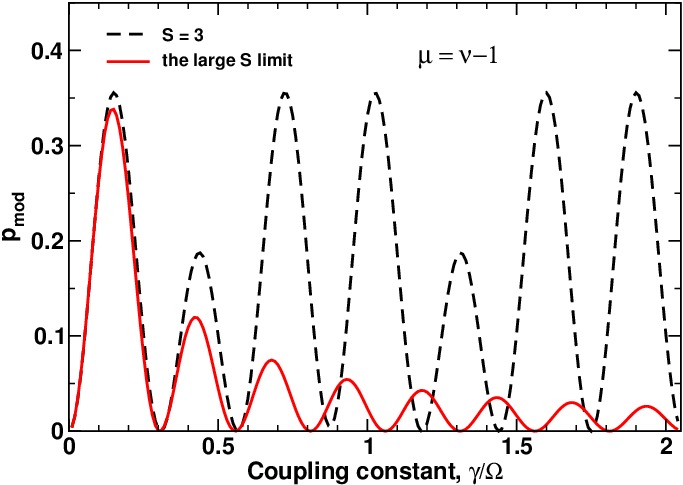}}
\label{subfig:01-nu-1}
}
\subfloat[]{
  \resizebox{70mm}{!}{\includegraphics*{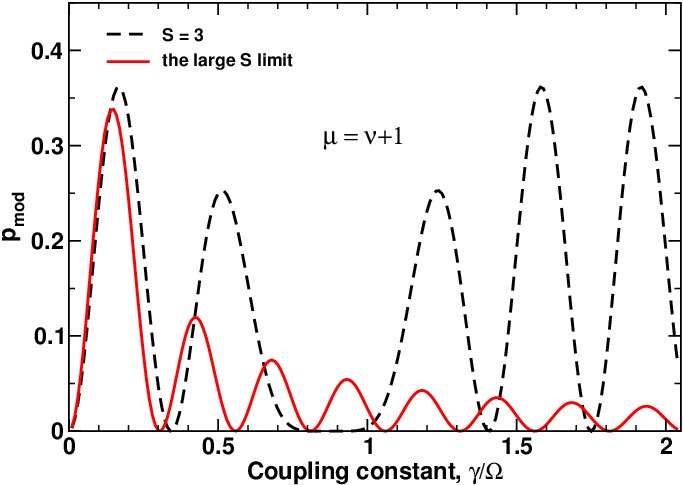}}
\label{subfig:21-nu-1}
}
\caption{%
(Color online)
The photon counting rate form-factor $p_{\ind{mod}}$ as a function of 
the coupling parameter $\gamma/\Omega$
for $\avr{N_\nu}(0)=\delta_{\nu,1}\avr{N_1}(0)$
at (a)~$\omega_f=\omega_0=\omega_{\ind{opt}}$
and (b)~$\omega_f=\omega_2=\omega_{\ind{opt}}+2\Omega$. 
}
\label{fig:nu-1}
\end{figure*}

When the pumped mode is not central
and $\nu\ne 0$,
the symmetry between the blue-detuned and red-detuned modes
with frequencies  
$\omega_{\nu}+ k\Omega$ and
$\omega_{\nu}- k\Omega$
appears to be broken
provided the number of interacting modes is finite.
Mathematically, the reason is the difference
between the magnitudes of the matrix elements
 $|U_{\nu+k, \nu}^S|$ and $|U_{\nu-k, \nu}^S|$.
By contrast, the symmetry retains
in the large $S$ limit where
$|U_{\nu\pm k, \nu}^S|\to |J_{\pm k}(g)|=|J_{|k|}(g)k|$.
The results computed
at $\nu=k=1$ are shown in Fig.~\ref{fig:nu-1}.
They clearly demonstrate pronounced asymmetry
between the modes with $\mu=\nu+1=2$ and $\mu=\nu-1=0$
that occurs at $S=3$, whereas 
the curves evaluated in the large $S$ limit
are clearly identical.

\subsection{Two-modulator transmission}
\label{subsec:two-modulator}

In conclusion of this section 
we briefly discuss how
our results can be extended to the important 
case where the input state
after being transformed by 
a  modulator of a sender
(Alice) is  transmitted through 
the optical fiber to a receiver (Bob) 
that sends the incoming state
through of a second modulator.
This is a simplified scheme representing 
the key elements used in 
frequency-coded setups~\cite{Merolla:prl:1999,Bloch:ol:2007}. 
We characterize the evolution operator of 
the system in terms of the matrix $\vc{M}$
[see Eq.~(\ref{eq:a-M})]
that enter the right hand side of Eq.~(\ref{eq:a-U}).
In our case, this matrix can be written as the product
of three matrices
\begin{align}
&
  \label{eq:M-prod}
  \vc{M}=\vc{M}_{2}\vc{M}_0\vc{M}_{1},
\\
&
  \label{eq:M-i}
M^{(0)}_{\mu\nu}=\delta_{\mu\nu}\e^{-i\Phi_\mu},
\quad
  M^{(i)}_{\mu\nu}=\e^{-i\Phi_{\mu\nu}^{(i)}} d_{\mu\nu}^{S}(\beta_i),
\end{align}
where 
the phase shift
$\Phi_{\mu}=\Phi_0+\mu\phi_0$
represents the effect of propagation
in the optical fiber
and the elements of
the Alice's(Bob's) modulator matrix,
$\vc{M}_1$ ($\vc{M}_2$),
are expressed in terms of the phase
given by 
\begin{align}
&
\label{eq:Phi-mu-nu}
\Phi_{\mu\nu}^{(i)}=\Phi_{00}^{(i)}+\mu
(\Omega_{\ind{MW}}T_i+\alpha_i+\phi_i)
+\nu (\pi+\alpha_i-\phi_i),
\end{align}
where $\Phi_{00}^{(i)}=\omega_{\ind{opt}}T_i$.
We assume that the only difference between 
otherwise identical modulators is
the phase of 
the microwave field,
$\phi_1=\phi_A$ and $\phi_2=\phi_B$,
that plays the role of the tuning parameter.
Other parameters of  the modulators are:
$T_1=T_2=T$, 
$\alpha_1=\alpha_2=\alpha_m$
and $\beta_1=\beta_2=\beta_m$.
Similar to Eq.~(\ref{eq:U-s}), 
we can use the relation
\begin{align}
&
\label{eq:rel-d}
\sum_{\mu'=-S}^{S}
d_{\mu\mu'}^{S}(\beta_m)
d_{\nu\mu'}^{S}(\beta_m)\e^{-i\mu'\phi_{AB}}=(-1)^{\nu}\e^{-i(\mu+\nu)\tilde\alpha}d_{\mu\nu}^{S}(\tilde\beta),
\\
&
\label{eq:gamma}
\phi_{AB}=\phi_{A}-\phi_B+\Delta,
\quad
\Delta=\phi_0+\Omega_{\ind{MW}}T+2\alpha_m
\end{align}
to derive the expression for the elements of the
matrix~(\ref{eq:M-prod})
in the final form:
\begin{align}
&
  \label{eq:M-2}
  M_{\mu\nu}=\e^{-i\Psi_{\mu\nu}} d_{\mu\nu}^{S}(\tilde{\beta}),
\\
&
\label{eq:Psi_mu_nu}
\Psi_{\mu\nu}=\psi_0+\mu(\Omega_{\ind{MW}}T+\alpha_m+\tilde{\alpha}+\phi_B)+
\nu (\pi+\alpha_m+\tilde{\alpha}-\phi_A),
\end{align}
where $\psi_0=\Phi_0+2 \omega_{\ind{opt}}T$.
The angles $\tilde{\alpha}$ and $\tilde\beta$
are determined by Eq.~(\ref{eq:talp-tbeta})
with the set of parameters $\{\Gamma T, \beta\}$
replaced by
$\{\phi_{AB}, \beta_m\}$.

In particular, from the suitably modified relation~\eqref{eq:c-tbeta} it follows that
$\cos\tilde\beta=1$ provided that $\cos\phi_{AB}=1$.
At these values of the tuning parameter $\phi_{AB}$,
the modulators compensate each other 
and $M_{\mu\nu}=\delta_{\mu\nu}$.
It implies that, in this regime, 
for the input light field without sidebands, 
no sidebands will be detected by Bob's photodetector.

Another limiting case is represented 
by the regime where 
the central optical mode is suppressed
after passing through Bob's modulator. 
This regime takes place when the condition
\begin{align}
  \label{eq:cond-zero-peak}
  d_{00}^{S}(\tilde\beta)\propto P_S(\cos\tilde\beta)=0,
\end{align}
where $P_S(x)$ is the Legendre polynomial,
is satisfied.

The intermode coupling
should be sufficiently strong,
$\gamma>\gamma_c$, 
for the condition~(\ref{eq:cond-zero-peak})
to be met.
To show this, we note that,
the value of $\cos\tilde\beta$
varies from unity 
to $\cos(2\beta_m)$
as the phase $\phi_{AB}$
changes from zero to $\pi$.
It implies that
the condition~(\ref{eq:cond-zero-peak}) 
cannot be fulfilled if
the value of $\cos(2\beta_m)$ 
is above
the largest root of the Legendre polynomial $P_S$
on the interval between zero and unity: $[0,1]$.
By making a simplifying assumption that
$\omega=0$ ($\Omega_{MW}=\Omega$) and $\beta=\pi/2$,
we find that $\beta_m=\Gamma T$
[$\Gamma$ is given by Eq.~(\ref{eq:Q-param})]. 
Then, for $T=2\pi/\Omega$ and $S=3$, 
the critical coupling ratio
$\gamma_c/\Omega$
can be numerically estimated to be at
about $0.0954$.
Interestingly, when the number of modes increases,
the critical coupling ratio approaches
the estimate $\gamma_c/\Omega\approx 0.0957$
obtained from 
the asymptotic form of the condition~(\ref{eq:cond-zero-peak}): 
$J_0(2 g)=0$, where $g$ is defined in Eq.~(\ref{eq:tbet-asymp}).

\section{Conclusions and discussion}
\label{sec:concl}
 
In this paper, we have formulated
a quantum multimode model of the electro-optic
modulator, where the intermode interaction
is induced by 
the microwave field via
the linear electro-optic effect (the Pockels effect).
This model is shown to be exactly solvable
when
the strength of coupling between
the interacting modes depends on the mode number
characterizing its detuning
from the central optical mode
and the operators 
[see Eqs.~(\ref{eq:J0}) and~(\ref{eq:Jpm})]
describing
the electro-optically induced interaction form the $su(2)$ Lie algebra
with the commutation relations given by Eq.~(\ref{eq:su2}).

Within the framework of the semiclassical approach
where the microwave field is treated as a classical signal
(the validity of this approximation is justified in Appendix~\ref{sec:quant-mw-field}),
we have used the analytical expressions for the quasienergy spectrum~(\ref{eq:E_quasi}) 
and the evolution operator~(\ref{eq:U-result})
in combination with the method of generalized Jordan mappings
(see Appendix~\ref{sec:jordan}) to describe
the temporal evolution of the photonic
annihilation (creation) operators in terms of the Wigner $D$~functions
[see Eqs.~(\ref{eq:a-U})--(\ref{eq:talp-tbeta})].
These results are then employed for
theoretical investigation into the effects of light modulation
on the photon counting rate. 
Based on the well-known
Mandel-Wolf
model of an idealized photodetector~\cite{Mandl:bk:1995}, 
we have found that the count rate computed as the one-electron
photodetection probability per unit time
can be written in the factorized form~(\ref{eq:P-2})
with the light modulation form-factor given by Eq.~(\ref{eq:p0-rel}).

Figures~\ref{fig:count-f-1}--~\ref{fig:nu-1}
present the numerical results for the counting rate form-factor
evaluated as a function of the frequency and the coupling constant.
In particular, the theoretical predictions
for the case where $S=3$
(the number of interacting modes equals $2S+1$)
are compared with
the large $S$ limit where $S$ increases indefinitely,
$S\to\infty$ (this limiting case is discussed in
Subsection~\ref{sec:results}.\ref{subsec:large-number-mode}).
It is found that the differences between these two cases
are negligible at small values of the coupling constant  
(see Fig.~\ref{fig:count-f-1}) and
become pronounced as the strength of intermode interaction
increases (see Figs.~\ref{fig:count-f-2}--~\ref{fig:count-g-2}).

In the large $S$ limit, coupling constant dependence of
the intensities of sidebands shows that 
the photons spread over available photonic states
leading to depletion of the pumped mode
(solid lines in Figs.~\ref{fig:count-g-1}--\ref{fig:nu-1}).
This is a consequence of
asymptotic behavior in the large $S$ limit
where, similar to the classical optics,
the effect of electro-optic light modulation
is shown to be determined by  
the modulating phase factor
given by Eq.~(\ref{eq:J-rel})
[see also Eq.~(\ref{eq:E-plus-asymp})].
 
By contrast, the intensities of sidebands
computed as a function of the coupling coefficient
at $S=3$
(dashed lines in Figs.~\ref{fig:count-g-1}--\ref{fig:nu-1})
appear to be nearly periodic.
Another interesting effect which
disappears in the large $S$ limit
can be described as the asymmetry in intensity
between the sidebands with the frequencies 
symmetrically arranged with respect
to the pumped mode 
(e.g. red shifted Stokes and blue shifted anti-Stokes modes).
As is illustrated in Fig.~\ref{fig:nu-1},
this asymmetry arises when the pumped mode
differs from the central one (the case of detuned pumping).

Analytical results are also employed to
describe the two-modulator transmission 
depending on 
the microwave phase difference.
We have studied the two important limiting regimes where either 
the modulators compensate each other
or
have a destructive effect on the central optical mode.
The latter is found to occur only if the intermode interaction
strength is sufficiently strond and exceeds 
its critical value. 

We now try to place our results into a more general physical context.
Generally, an exactly solvable model 
where the electro-optic modulator is viewed 
as a multiport device can be employed as 
a theoretical tool for investigation into numerous effects 
coming from the complicated quantum dynamics of
multimode systems.  
In addition, this model
deals with parametric processes
that play important part
in the so-called 
resonator optomechanics~\cite{Aspelmeyer:bk:2014,Aspel:rmp:2014} 
representing a 
new branch of quantum information science
that rapidly evolves at the interface of  
the nanophysics and the quantum theory of light. 
Making progress in studies of 
the Casimir effect, new protocols of quantum communication, 
quantum computing and quantum memory
will require further insight into the theory of such
parametric processes.

Mathematically, we have demonstrated in
Appendix~\ref{sec:quant-mw-field}
that it is feasible to 
apply the methods of 
polynomially deformed algebras~\cite{Vadeiko:pra:2003}
to extend our considerations to the case of
quantized microwave field.
This case, however, requires a more comprehensive study
which is beyond the scope of this paper.
On the other hand, our approach provides a useful
tool for investigation of high-frequency light modulation
in liquid crystal modulators driven by 
the orientational Kerr
effect~\cite{Kiselev:pre:2013,Kiselev:pre:2:2014,Kiselev:ol:2014,Kiselev:pre:2015}.
In particular, the model
can be generalized to take into account
effects of non-trivial 
polarization dependent quantum dynamics.
This work is now in progress.

We acknowledge partial financial support
from the Government of the Russian Federation (Grant No.
074-U01).

\appendix

\titleformat{\section}
{\large\sffamily\bfseries}
{APPENDIX \thesection:}
{0.5em}
{\MakeUppercase{#1}}
[]

\setcounter{equation}{0}
\renewcommand{\theequation}{\thesection{\arabic{equation}}}

\section{Jordan mapping technique}
\label{sec:jordan}

These mappings are defined as follows
\begin{align}
  \label{eq:Jordan-map}
  \mvc{J}_{\alpha}\mapsto
J_{\alpha}=\sum_{\nu,\mu=-S}^{S}\hcnj{a}_{\nu}J_{\nu\mu}^{(\alpha)}
a_{\mu}
\equiv
\hcnj{\vc{a}}\mvc{J}_{\alpha} \vc{a},
\end{align}
where $\mvc{J}_{\alpha}$ 
is the $(2S+1)\times(2S+1)$ matrix.
The elements $J_{\nu\mu}^{(\alpha)}$
($\equiv [\mvc{J}_{\alpha}]_{\nu\mu}$) 
of the matrices $\mvc{J}_{\alpha}$ 
with $\alpha\in\{0,\pm\}$ are given by
\begin{align}
  \label{eq:Jnumu}
  J_{\nu\mu}^{(\pm)}=\sqrt{(S\mp\mu)(S\pm\mu+1)}\delta_{\nu\mu\pm 1},
\quad
  J_{\nu\mu}^{(0)}=\mu\delta_{\nu\mu},
\end{align}
where $\delta_{\nu\mu}$ is the Kronecker symbol.
Using the standard bosonic commutation relations
\begin{align}
  \label{eq:boson-comm}
  [a_{\nu},\hcnj{a}_{\mu}]=\delta_{\nu\mu},
\quad
  [\hcnj{a}_{\nu},\hcnj{a}_{\mu}]=[a_{\nu},a_{\mu}]=0
\end{align}
it is not difficult to check the key useful property
of the Jordan construction:
\begin{align}
  \label{eq:Jord-comm}
  [J_{\alpha},J_{\beta}]=\hcnj{\vc{a}}[\mvc{J}_{\alpha},\mvc{J}_{\beta}] \vc{a}.
\end{align}
The result~(\ref{eq:su2})
follows because the matrices
$\mvc{J}_{\pm}$ and $\mvc{J}_{0}$
with the elements given in Eq.~(\ref{eq:Jnumu})
satisfy the commutation relations for $su(2)$ algebra.
Another useful relation can be derived for
the Baker-Campbell-Haussdorf formula
\begin{align}
  \label{eq:BCH}
  \exp(i\beta J_{\alpha}) a_{\mu} \exp(-i\beta J_{\alpha})
=\sum_{k=0}^{\infty}\frac{i^k\beta^k}{k!}[J_{\alpha},a_{\mu}]_{(k)},
\end{align}
where $[J_{\alpha},a_{\mu}]_{(k)}$ stands for the multiple commutator
\begin{align}
&
  \label{eq:mult-comm}
  [J_{\alpha},a_{\mu}]_{(k)}=[J_{\alpha},[J_{\alpha},a_{\mu}]_{(k-1)}],
\notag
\\
&
 [J_{\alpha},a_{\mu}]_{(1)}= [J_{\alpha},a_{\mu}],
\quad
 [J_{\alpha},a_{\mu}]_{(0)}=a_{\mu}.
\end{align}
From Eqs.~(\ref{eq:Jordan-map})
and~(\ref{eq:boson-comm}) we have
\begin{align}
  \label{eq:J-a-a-comm}
  [J_{\alpha},a_{\mu}]=
-\sum_{\nu=-S}^{S} J_{\mu\nu}^{(\alpha)} a_{\nu},
\end{align}
and formula~(\ref{eq:BCH}) can be recast
into the final form 
\begin{align}
  \label{eq:BCH-J}
  \exp(i\beta J_{\alpha}) a_{\mu} \exp(-i\beta J_{\alpha})
=\sum_{\nu=-S}^{S} [\exp(-i\beta \mvc{J}_{\alpha})]_{\mu\nu}a_{\nu}.
\end{align}

An important consequence of Eq.~(\ref{eq:BCH-J}) is 
the identity
\begin{align}
\label{eq:a-R}
  \e^{i\gamma J_z}\e^{i\beta J_y}\e^{i\alpha J_z} a_{\mu}
  \e^{-i\alpha J_z}\e^{-i\beta J_y}\e^{-i\gamma J_z}=
&
\sum_{\nu=-S}^{S}[\e^{-i\alpha \mvc{J}_z}\e^{-i\beta \mvc{J}_y}\e^{-i\gamma
  \mvc{J}_z}]_{\mu\nu} a_{\nu}
\notag
\\
&
=\sum_{\nu=-S}^{S} D_{\mu\nu}^{S}(\alpha,\beta,\gamma) a_{\nu}
\end{align}
for the rotated annihilation operator
expressed in terms of
the Wigner $D$ functions:
$D_{\mu\nu}^{S}(\alpha,\beta,\gamma)=\exp[-i\mu\alpha]d_{\mu\nu}^{S}(\beta)\exp[-i\nu\gamma]$ 
that, for the irreducible representation of the rotation group
with the angular number $S$,
give the elements of the rotation matrix parametrized by the three
Euler angles~\cite{Biedenharn:bk:1981,Varshalovich:bk:1988}:
$\alpha$, $\beta$ and $\gamma$.

We conclude this section with details
on derivation of the expression
for the matrix elements
of the operator $\mvc{U}_S(t)$
given in Eq.~(\ref{eq:U-s}).
This operator can be written in the form
\begin{align}
  \label{eq:R_m_t}
  \mvc{U}_S(t)=\e^{-i\beta \mvc{J}_y}\e^{-i\Gamma t \mvc{J}_z}
\e^{i\beta \mvc{J}_y}=
\e^{-i\Gamma t (\sin\beta\mvc{J}_x+\cos\beta\mvc{J}_z)}.
\end{align}
More generally, we consider the rotation operator
\begin{align}
  \label{eq:R-gen-m-t}
  R(\psi,\uvc{m})=\exp[-i \psi \sca{\uvc{m}}{\mvc{J}}],
\end{align}
where $\mvc{J}=(J_x,J_y,J_z)$,
$\psi=\Gamma t$ 
and 
$\uvc{m}=(m_x,m_y,m_z)\equiv (m_1,m_2,m_3)=(\sin\beta,0,\cos\beta)$
is the unit vector directed along the rotation axis.
Equation~(\ref{eq:R-gen-m-t})
defines rotation about the rotation axis $\uvc{m}$
by the rotation angle $\psi=\Gamma t$.
Alternatively, this rotation can be
parametrized by the Euler angles
as follows
\begin{align}
  \label{eq:R-abg}
  R(\psi,\uvc{m})=R(\tilde\alpha,\tilde\beta,\tilde\gamma)=
  \e^{-i\tilde\alpha J_z}\e^{-i\tilde\beta J_y}\e^{-i\tilde\gamma J_z}.
\end{align}
Our task is to express the Euler angles
$\tilde\alpha$, $\tilde\beta$ and $\tilde\gamma$
in terms of the rotation angle $\psi=\Gamma t$ and 
the angle of the rotation axis $\beta$.
To this end, we begin with the relations
\begin{align}
&
  \label{eq:rel-m}
  R(\psi,\uvc{m})\sca{\vc{n}}{\mvc{J}}\hcnj{R}(\psi,\uvc{m})=
\sca{\mvc{R}(\psi,\uvc{m})\vc{n}}{\mvc{J}},
\\
&
  \label{eq:rel-abg}
R(\tilde\alpha,\tilde\beta,\tilde\gamma)\sca{\vc{n}}{\mvc{J}}\hcnj{R}(\tilde\alpha,\tilde\beta,\tilde\gamma)=
\sca{\mvc{R}(\tilde\alpha,\tilde\beta,\tilde\gamma)\vc{n}}{\mvc{J}},
\end{align}
where $\mvc{R}(\psi,\uvc{m})$
and $\mvc{R}(\tilde\alpha,\tilde\beta,\tilde\gamma)$
are the $3\times 3$ rotation matrices,
that hold for arbitrary vector $\vc{n}$.
\begin{align}
  \label{eq:R3-m}
  \mvc{R}(\psi,\uvc{m})=
\mvc{I}_3\cos\psi+\uvc{m}\otimes\uvc{m}
(1-\cos\psi)+\mvc{M}\sin\psi,
\end{align}
where $\mvc{I}_3$ is the $3\times 3$ identity matrix
and $\mvc{M}$ is the antisymmetric matrix
with the elements 
$\displaystyle M_{ij}=-\sum_{k=1}^3\epsilon_{ijk}m_k$
defined using the unit vector $\uvc{m}$
and the antisymmetric tensor $\epsilon_{ijk}$
($\epsilon_{123}=1$). In our case, we have
\begin{align}
  \label{eq:M}
  \mvc{M}=
  \begin{pmatrix}
    0&\cos\beta&0\\
\cos\beta&0&-\sin\beta\\
0&\sin\beta&0
  \end{pmatrix}.
\end{align}
From the other hand, the rotation matrix
$\mvc{R}(\tilde\alpha,\tilde\beta,\tilde\gamma)$
is given by
\begin{align}
  \label{eq:R3-abg}
  \mvc{R}(\tilde\alpha,\tilde\beta,\tilde\gamma)=
  \mvc{R}_z(\tilde\alpha) \mvc{R}_y(\tilde\beta)
  \mvc{R}_z(\tilde\gamma)
\end{align}
a product of the rotation matrices of the form:
\begin{align}
  \label{eq:Rxy}
  \mvc{R}_z(\tilde\alpha)=
  \begin{pmatrix}
    \cos\tilde\alpha&-\sin\tilde\alpha&0\\
\sin\tilde\alpha&\cos\tilde\alpha&0\\
0&0&1
  \end{pmatrix},
\:
  \mvc{R}_y(\tilde\beta)=
  \begin{pmatrix}
    \cos\tilde\beta&0&\sin\tilde\beta\\
0&1&0\\
-\sin\tilde\beta&0&\cos\tilde\beta
  \end{pmatrix}.
\end{align}
The relations linking different parametrizations
can now be obtained from the condition:
\begin{align}
  \label{eq:RR-eq}
\mvc{R}(\psi,\uvc{m})=
  \mvc{R}(\tilde\alpha,\tilde\beta,\tilde\gamma)\equiv\mvc{R}.
\end{align}

Since, for the matrix $\mvc{R}(\psi,\uvc{m})$, 
$R_{13}=R_{31}$, $R_{21}=-R_{12}$ and
$R_{23}=-R_{32}$, we have
\begin{align}
  \label{eq:tgamma}
  \tilde\gamma=\tilde\alpha+\pi
\end{align}
and the condition~(\ref{eq:RR-eq})
gives the following relations:
\begin{subequations}
\label{eq:Rij}
\begin{align}
&
  \label{eq:R12}
  -\sin(2\tilde\alpha)(1+\cos\tilde\beta)=2\sin\beta\sin\psi=R_{21},
\\
&
  \label{eq:R11-R22}
  -\cos(2\tilde\alpha)(1+\cos\tilde\beta)=\sin^2\beta+(1+\cos^2\beta)\cos\psi=
R_{11}+R_{22},
\\
&
  \label{eq:R33}
  \cos\tilde\beta=\cos^2\beta+\sin^2\beta\cos\psi=R_{33},
\\
&
  \label{eq:R13}
  \cos\tilde\alpha\sin\tilde\beta=\sin\beta\cos\beta(1-\cos\psi)=R_{13},
\\
&
  \label{eq:R23}
  \sin\tilde\alpha\sin\tilde\beta=-\sin\beta\sin\psi=R_{23}.
\end{align}
\end{subequations}

From Eqs.~(\ref{eq:R12}) and~(\ref{eq:R11-R22}),
we derive the expression for the angle $\tilde\alpha$
given in Eq.~(\ref{eq:talp} 
whereas the angle $\tilde\beta$ is described by 
formulas~(\ref{eq:c-tbeta}) and~(\ref{eq:s-tbeta})
that can be easily obtained from
Eqs.~(\ref{eq:R33})--(\ref{eq:R23}).

Our concluding remarks concern two special cases where either
$\sin\beta=0$ or $\cos\beta=0$.
When $\sin\beta=0$ and  $\cos\beta=\pm 1$,
the operator~(\ref{eq:R-gen-m-t}) describes rotations
about the $z$ axis by the angle $\pm\psi$ and the angles
$\tilde\alpha$, $\tilde\beta$ and $\tilde\gamma$ are given by
\begin{align}
  \label{eq:sin-0}
  \tilde\beta=0,
\quad
\tilde\alpha+\tilde\gamma=\pm\psi.
\end{align}
At $\cos\beta=0$ and  $\sin\beta=\pm 1$, 
the rotation axis is parallel to the $x$ axis
and we have
\begin{align}
  \label{eq:cos-0}
  \tilde\beta=\pm\psi,
\quad
\tilde\gamma=-\tilde\alpha=\pi/2.
\end{align}

\setcounter{equation}{0}
\renewcommand{\theequation}{\thesection{\arabic{equation}}}

\section{Quantized microwave field and polynomially deformed algebras}
\label{sec:quant-mw-field}

In the model with the Hamiltonian~(\ref{eq:Hamilt-s}) 
the microwave field is treated as a classical field
characterized by the c-number amplitude $B$. 
In this appendix we briefly discuss how this model can be extended
to the case where, similar to the optical modes, 
the microwave field is quantized.
In our analysis we employ the technique of polynomially deformed
algebras to study
applicability of the semiclassical approach.

For full quantum description of the modes, 
we begin with the
Hamiltonian~(\ref{eq:Hamilt-0})
rewritten as follows
\begin{align}
H/\hbar = \Omega_{MW} N_b + \omega_{\ind{opt}} N + 
\Omega J_z + \frac{2 \gamma_0}{2S+1}\left({J_{+} b + J_{-}\hcnj{b}}
\right),
\label{eq:Hq}  
\end{align}
where
$N_b=\hcnj{b} b$,
the operators $J_z$ and $J_{\pm}$ 
given by Eqs.~(\ref{eq:A0})--(\ref{eq:Jpm})
meet the commutation relations
for generators of $su(2)$ algebra~(\ref{eq:su2}),
whereas the creation and annihilation operators
of the microwave mode,
$\hcnj{b}$ and $b$, obey the commutation relation 
of the Heisenberg-Weyl algebra:
$[b,\hcnj{b}]=1$. 

A set of operators that commute with
the Hamiltonian~(\ref{eq:Hq}) contains three operators: 
(a)~the operator of the total photon number 
for the optical modes $N$ given in Eq.~(\ref{eq:A0});
(b)~the Casimir operator of $su(2)$ algebra $J^2$ 
given by Eq.~(\ref{eq:J2}); and 
(c)~the additional operator
$R=N_b+J_z$ related to the non-negative 
excitation number operator
$M=N_b+J_z+j I$, where
$I$ is the identity operator and $j$
is the angular momentum quantum number
[$j(j+1)$ is the eigenvalue of $J^2$].

The Fock states for the model under consideration
are represented by a direct product of the microwave and optical Fock states:
$\ket{n_b}_b\otimes\ket{\psi}_a$,
where $n_b$ is the photon number of the microwave mode.
The Fock space
can be conveniently divided into subspaces  $\mathcal{F}_{n,\,m,\,j}$
classified by the quantum numbers
$m$, $n$ and $j$,
where $m$, $n$ and $j(j+1)$ are
the eigenvalues of the operators $M$, $N$ and $J^2$,
respectively.
The basis  of  $\mathcal{F}_{n,\,m,\,j}$
can be formed from the eigenstates of
the operator $J_z$ 
\begin{align}
  \label{eq:ket-mnj}
  \ket{m,n,j,m_z}=\ket{m-m_z-j}_b\otimes\ket{n,j,m_z}_a,
\end{align}
where $-j\le m_z\le \min\{j,m-j\}$
is the azimuthal quantum number
[the microwave photon number
$n_b=m-m_z-j$ is a nonnegative integer]
and $J_z\ket{m,n,j,m_z}=m_z\ket{m,n,j,m_z}$.
Clearly, the quantum
numbers $m$ and $j$ determine
dimension of $\mathcal{F}_{n,\,m,\,j}$. 
At $m\ge 2j$, the quantum number $m_z$ 
is ranged from $-j$ to $j$ and
$\dim\mathcal{F}_{n,\,m,\,j}=2j+1$.
In the opposite case with $m< 2j$,
we have $-j\le m_z\le m-j$ and
$\dim\mathcal{F}_{n,\,m,\,j}=m+1$.

In  the subspace $\mathcal{F}_{n,\,m,\,j}$,
the Hamiltonian~(\ref{eq:Hq}) 
is reduced to the following form:
 \begin{align}
H/\hbar = n\,\omega_{\ind{opt}} + r \tilde{\Omega} -\omega M_0 + 
\frac{2 \gamma_0}{2S+1}
\left(M_{+}  + M_{-}\right), 
\label{eq:Hq2}
\end{align}
where 
$\tilde{\Omega}=(\Omega+\Omega_{MW})/2$,
$r=m-j$ is the eigenvalue of the operator $R=N_b+J_z$
and
the operators
$M_0$ and $M_{\pm}$ are given by
\begin{align}
M_{-} = b J_{+}, \quad M_{+} = \hcnj{b} J_{-}, \quad M_{0} = \frac{N_b - J_z}{2}.
\label{eq:Genq}
\end{align}

We can now closely follow 
the line of reasoning described in Ref.~\cite{Vadeiko:pra:2003}
and apply the methods of deformed (quantum) Lie algebras
to solve the spectral problem
for the Hamiltonian~(\ref{eq:Hq2}). 
For this purpose, we note that
the operators~(\ref{eq:Genq})
can be regarded as the generators
of polynomial algebra of excitations (PAE).
This algebra is generally defined through 
the algebraic relations:    
\begin{align}
\left[ M_0, M_{\pm}\right] = \pm M_{\pm}, \quad 
M_{+} M_{-} = p_{\kappa}(M_0), 
\label{eq:crq}
\end{align}
where $p_{\kappa}(q)$ is the structure
polynomial of degree $\kappa$ 
characterizing PAE of order $\kappa$. 
In our case, we have
\begin{align}
&
\label{eq:MpMm}
M_{+}M_{-} = N_b \left( {J^2 - J_z^2 - J_z}\right) = 
p_3(M_0), 
\\
&
\label{eq:p3}
 p_3(q) = -(q-q_1)(q-q_2)(q-q_3),     
\end{align}
where the roots of the polynomial $p_3$ are given by
\begin{align}
\label{eq:roots}
q_1 = \frac{j-m}{2}, \quad q_2 = \frac{m-3j}{2}, \quad q_3 = \frac{m+j}{2} + 1.
\end{align}
The structure polynomial~(\ref{eq:p3}) defines PAE of third order
that will be denoted by $\mathcal{M}_{m,j}$. 
Since $m\ge 0$, the largest root is $q_3$,  whereas relation between $q_1$ and
$q_2$ depends on the values of $m$ and $j$:
$q_1 > q_2$ at $m < 2j$ and
$q_2 > q_1$ at $m > 2j$.
When differences
$d_1 = q_3 - q_1 = m + 1$
and $d_2 = q_3 - q_2 = 2j + 1$ 
are natural numbers, $d_i\in\mathbb{N}$,
algebra $\mathcal{M}_{m,j}$ is known to have 
finite-dimensional self-adjoint representations
that correspond to the positive spectrum of $p_3(M_0)$.

When $m>2j$ [$r>j$], the finite-dimensional irreducible representation of
$\mathcal{M}_{m,j}$ will be referred to as the \textit{high-excitation zone}.
Its dimension equals $2j+1$ and the corresponding spectrum of $p_3(M_0)$
is ranged from $q_2$ to $q_3$.
In the opposite case with $m<2j$ [$r<j$],
the positive part of the spectrum lies in the interval
$[q_1,q_3]$ and the dimension of 
the representation~---~
the so-called \textit{low-excitation zone}~---~is equal to $m+1$.

In the method of Ref.~\cite{Vadeiko:pra:2003},
the technique of polynomially deformed algebra
is used to construct 
the transformations that map one polynomial
algebra of operators onto another. 
More specifically, 
the representation of algebra
$\mathcal{M}_{m,j}$
with the generators
$\{M_0,M_{+},M_{-}\}$
is related to a simpler algebra of second order
with the generators $\{S_0,S_{+},S_{-}\}$
that meet the commutation relations
of $su(2)$ algebra~(\ref{eq:su2})
and its irreducible representation
is characterized by
the angular quantum number $s$.
The number $s$ is fixed
by the requirement for
two representations to be of the same dimension.
Mathematical details 
on the method
and a more accurate 
formulation of the key statements
can be found in Ref.~\cite{Vadeiko:pra:2003}.

\subsection{High-excitation zone}
\label{subsec:high-excitation-zone}

First we consider the important case
of the high-excitation zone, where
$s=j$ and the operators
$\{M_0,M_{+},M_{-}\}$
are expressed in terms of $\{S_0,S_{+},S_{-}\}$
as follows~\cite{Vadeiko:pra:2003}
\begin{align}
&
\label{eq:M-S-s}
  M_0 = \frac{r}{2} - S_0, 
\quad M_{+} = \sqrt{r - S_0}\, S_-, 
\notag
\\
& 
M_{-}= \hcnj{[M_{+}]}=S_+\, 
  \sqrt{r - S_0},
\end{align}
where $r=m-j$.
It is also not difficult to obtain the relations 
\begin{align}
\label{eq:S-M-s}
S_0 = J_z, 
\quad 
S_+ = \frac{1}{\sqrt{N_b+1}} a J_+, 
\quad 
S_- = J_- \hcnj{a} \frac{1}{\sqrt{N_b+1}}
\end{align}
linking
$\{S_0,S_{+},S_{-}\}$
and the operators that enter the Hamiltonian~(\ref{eq:Hq}).

We can now substitute relations~(\ref{eq:M-S-s})
into Eq.~(\ref{eq:Hq2}) to obtain the Hamiltonian
expressed in terms of the operators
$\{S_0,S_{+},S_{-}\}$.
In the zero-order approximation, we have 
\begin{align}
\label{eq:zero-s}
M_0 = r/2 - S_0, 
\quad M_{\pm} \approx 
\sqrt{r + 1/2}\,S_{\mp},  
\end{align}
so that
the approximate structure polynomial
\begin{align}
\label{eq:p2-strong}
p_2^{(s)}(M_0) = (r + 1/2) S_{-} S_{+} = 
- (r+1/2)(M_0 - q_2 )(M_0 - q_3)  
\end{align}
is quadratic.
The corresponding zero-order Hamiltonian is given by
\begin{align}
\label{eq:H-0-s}
H_{0}^{(s)}/\hbar =  n\,\omega_{\ind{opt}}  + r \Omega_{MW} +
\omega S_0 + \frac{2 \gamma_0}{2S+1} \sqrt{r+1/2} \left({S_+ + S_-}
\right).
\end{align}
A comparison between $H_{0}^{(s)}$
and the quasienergy operator for 
the semiclassical model~(\ref{eq:Q})
shows that these operators are similar in algebraic structure.
In particular, 
similar to formula~(\ref{eq:Q}),
the quantum number $j$
that determines the dimension of the representation
does not enter the expression for $H_{0}^{(s)}$.
So, when $\gamma$ is replaced by
$\gamma_0\sqrt{r+1/2}$, 
the spectra of these operators are identical 
up to the additive constant. 
We thus may conclude that 
the zero-order approximation
for the high-excitation zone of the model
with quantized microwave field
reproduces the results of semiclassical approach.
Note that the condition $r>n_{\ind{max}}S\equiv j_{\ind{max}}$
ensures applicability of the semiclassical approximation
for the Fock states of the optical modes whose
total photon numbers are below $n_{\ind{max}}$.

\subsection{Low-excitation zone}
\label{subsec:low-excitation-zone}

In conclusion, we briefly review the results for
the low-excitation zone where $m< 2j$.
The dimension of the representation is
now equal to $m+1$, so that $s=m/2$. 
The corresponding positive part of $p_3(M_0)$ spectrum
is ranged from $q_1=(j-m)/2$ to $q_3=(m+j)/2+1$
and relations linking
$\{M_0,M_{+},M_{-}\}$
and $\{S_0,S_{+},S_{-}\}$
are given by
\begin{align}
&
\label{eq:M-S-w}
M_0 = \frac{j}{2} - S_0, 
\notag
\\
& 
M_+ = \sqrt{ 2j - m/2 - S_0}\,S_{-}, 
\quad 
M_- = S_+\sqrt{2j - m/2 - S_0},
\\
&
\label{eq:S-M-w}
S_0 = \frac{m}{2}-N_b, 
\quad 
S_+ = J_+ \frac{1}{\sqrt{j-J_0}}a,
\quad 
S_- = \hcnj{a} \frac{1}{\sqrt{j-J_0}} J_-.
\end{align}

In the zero-order approximation,
the operators~(\ref{eq:M-S-w}) are simplified
as follows
\begin{align}
\label{eq:zero-w}
M_0 = -S_0 + \frac{j}{2}, 
\quad 
M_{\pm} \approx \sqrt{(1-m)/2+2j}\, S_{\mp} 
\end{align}
and
the approximate structure polynomial is given by
\begin{align}
\label{eq:p2-weak}
p_2^{(w)}(M_0) = -[(1-m)/2+2j](M_0-q_1)(M_0-q_3).    
\end{align}
Finally, substituting relations~(\ref{eq:zero-w})
into formula~(\ref{eq:Hq2}) yields 
the expression for the zero-order Hamiltonian
in the low-excitation zone
\begin{align}
&
\label{eq:H-0-w}
H_{0}^{(w)}/\hbar =n\, \omega_{\ind{opt}} + \frac{m}{2} \Omega_{MW} 
+\frac{r}{2} \Omega + \omega S_0 
\notag
\\
&
+ 
\frac{2 \gamma_0}{2S+1} \sqrt{(1-m)/2+2j} \left({S_+ + S_-} \right).
\end{align}
In contrast to the case of the high-excitation zone,
the parameters of the Hamiltonian~(\ref{eq:H-0-w})
and the dimension of the representation
both depend on the quantum numbers $r$ and $j$.
So, the semiclassical approximation
breaks down in the low-excitation zone
and quantum effects become essential
for description of this regime even in the zero-order approximation.


\end{document}